\documentstyle[psfig,graphics]{mn2e}
\title [Galactic model parameters]
{The investigation of ELAIS field by Vega photometry: 
Absolute - magnitude dependent on the Galactic model parameters
}
\author[Bilir et al.]
       {S.~Bilir,$^1$ \thanks{E-mail: sbilir@istanbul.edu.tr}
        S.~Karaali$^1$, and G.~Gilmore$^{2}$\\
  $^1$Istanbul University Science Faculty, Department of Astronomy and Space
      Sciences, 34119, University-Istanbul, Turkey\\
  $^2$ Institute of Astronomy, Madingley Road, Cambridge, CB3 OHA, UK\\}

\pagerange{\pageref{firstpage}--\pageref{lastpage}}
\begin{document}

\maketitle

\label{firstpage}

\begin{abstract}

We estimate the density laws of the Galactic stellar populations as a function of 
absolute magnitude in a near-polar Galactic field. The density laws are 
determined by the direct fit to photometric parallaxes from Vega photometry in the 
ELAIS ($\alpha=16^{h}10^{m}00^{s}, \delta= +54^{o}30^{'}00{''}$; $l=84^{o}.27$, 
$b=+44^{o}.90$; 6.571 deg$^{2}$; epoch 2000) field both independently for each 
population and simultaneously for all stellar populations. Stars have been 
separated into different populations based on their spatial location. The thick 
disc and halo best fit by an exponential. However, the thin disc best fits by 
using a sech$^{2}$ law for stars at faint absolute magnitudes, $10<M(g^{'})\leq11$,    
$11<M(g^{'})\leq12$ and $12<M(g^{'})\leq13$, whereas an exponential law for 
stars at relatively bright absolute magnitudes, $5< M(g^{'})\leq 6$, 
$6<M(g^{'})\leq7$, $7<M(g^{'})\leq8$, $8<M(g^{'})\leq9$ and $9<M(g^{'})\leq10$. 
The scaleheights for the sech$^{2}$ density laws are the equivalent exponential 
scaleheights. Galactic model parameters are absolute magnitude dependent: The 
scaleheight for thin disc decreases monotonically from stars at bright absolute 
magnitudes $[M(g^{'})=5]$ to stars at faint absolute magnitudes $[M(g^{'})=13]$ in 
the range 363-163 pc, except the minimum H=211 pc at $9<M(g^{'})\leq10$ where 
sech density law fits better. Its local density is flat at bright absolute 
magnitudes but it increases at faint absolute magnitudes. For thick disc, the 
scaleheight is flat within the uncertainties. The local space density of thick 
disc relative to the local space density for the thin disc is almost flat at 
absolute magnitude intervals $5<M(g^{'})\leq6$ and $6<M(g^{'})\leq7$, 7.59 and 
7.41 per cent respectively, whereas it decreases down to 3.31 per cent at absolute 
magnitude interval $7<M(g^{'})\leq8$. The axial ratio for the halo is $\kappa$
=0.60, 0.73 and 0.78 for the absolute magnitude intervals $4<M(g^{'})\leq5$,   
$5<M(g^{'})\leq6$ and $6<M(g^{'})\leq7$ respectively, and its local space density 
relative to the local space density for the thin disc is 0.06 and 0.04 per cent 
for the intervals $5<M(g^{'})\leq6$, and $6<M(g^{'})\leq7$ respectively (the local 
space density relative to the thin disc could not be derived for the absolute 
magnitude interval $4<M(g^{'})\leq5$ due to the lack of the local space 
density for thin disc for this interval). The simultaneous fit of all three 
stellar populations agrees within uncertainties with the most recent values in 
the literature. Also, each parameter is close to one of the corresponding 
parameters estimated for different absolute magnitude intervals in this work with 
one exception however; i.e. the scaleheight for thick disc is relatively small 
and its error is rather large ($H=760^{+62}_{-55}$ pc).   
\end{abstract}


\begin{table*}
\center
\caption{Previous Galactic models. Symbols: TN denotes the thin disc, TK denotes 
the thick disc, S denotes the spheroid (halo), $R_{e}$ is the effective radius and 
$\kappa$ is the axes ratio. The figures in the parentheses for Siegel et al.\ 
(2002) are the corrected values for binarism. The asterisk denotes the 
power-law index replacing $R_{e}$.}
{\scriptsize
\begin{tabular}{lllllllll}
\hline
H (TN)& h (TN)& n (TK)& H (TK) & h (TK) & n (S) & $R_{e}$ (S) & $\kappa$&  Reference \\
(pc) & (kpc) & &  (kpc) & (kpc) & & (kpc) & & \\
\hline
310-325 & --- & 0.0125-0.025 &  1.92-2.39 & --- & --- & --- & --- & Yoshii\ (1982) \\
300 & --- & 0.02 & 1.45 & --- & 0.0020 & 3.0 & 0.85 & Gilmore \& Reid\ (1983)\\
325 & --- & 0.02 & 1.3 &  --- & 0.0020 & 3.0 & 0.85 & Gilmore\ (1984) \\
280 & --- & 0.0028 &  1.9 & --- &0.0012 &--- & --- & Tritton \& Morton\ (1984) \\
200-475 & --- & 0.016 &  1.18-2.21 & --- & 0.0016 & --- & 0.80 & 
Robin \& Cr\'{e}z\'{e}\ (1986) \\
300 & --- & 0.02 & 1.0 & --- & 0.0010 & --- & 0.85 & del Rio \& Fenkart\ (1987) \\
285 & --- & 0.015 &1.3-1.5 & --- & 0.0020 & 2.36 & Flat & Fenkart et al.\ (1987) \\
325 & --- & 0.0224 & 0.95 & --- & 0.0010 & 2.9 & 0.90 & Yoshii, Ishida \& Stobie\ (1987) \\
249 & --- & 0.041 & 1.0 & --- & 0.0020 & 3.0 & 0.85 & Kuijken \& Gilmore\ (1989) \\
350 & 3.8 & 0.019 & 0.9 & 3.8 & 0.0011 & 2.7 & 0.84 & Yamagata \& Yoshii\ (1992) \\
290 & --- & --- & 0.86 & --- & --- & 4.0 & --- & von Hippel \& Bothun\ (1993) \\
325 & --- & 0.0225 & 1.5 & --- & 0.0015 & 3.5 & 0.80 & Reid \& Majewski\ (1993) \\
325 & 3.2 & 0.019 & 0.98 & 4.3 & 0.0024 & 3.3 & 0.48 & Larsen\ (1996) \\
250-270 & 2.5 & 0.056 & 0.76 & 2.8 & 0.0015 & 2.44-2.75* &  0.60-0.85 & 
Robin et al.\ (1996); Robin, Reyl\'{e} \& Cr\'{e}z\'{e}(2000) \\
290 &4.0 & 0.059 & 0.91 & 3.0 & 0.0005 & 2.69 & 0.84 & Buser, Rong \& Karaali\ (1998, 1999) \\
240 &2.5 & 0.061 & 0.79 & 2.8 & --- & --- & 0.60-0.85 & Ojha et al.\ (1999) \\
330 &2.25& 0.065-0.13 &  0.58-0.75 & 3.5 & 0.0013 & --- & 0.55 & 
Chen et al.\ (2001) \\
280(350) & 2-2.5&  0.06-0.10 & 0.7-1.0 (0.9-1.2) & 3-4 & 0.0015 & --- & 
0.50-0.70 & Siegel et al.\ (2002)\\
320 & --- & 0.07 & 0.64 & --- & 0.0013 & --- & 0.58 & Du et al.\ (2003) \\
265-495 & --- & 0.052-0.095 & 0.80-0.97 & --- & 0.0002-0.0015 & --- &  0.70 & 
Karaali, Bilir \& Hamzao\u glu\ (2004) \\
\hline
\end{tabular}
}  
\end{table*} 

\section{Introduction}

Galactic models have a long history. Bahcall \& Soneira\ (1980) fitted their 
observations with a double component Galactic model, namely disc and halo, 
whereas Gilmore \& Reid\ (1983) could succeed to fit their observations with 
a Galactic model only by introducing a third component, i.e. thick disc. It 
should be noted that the third component was a rediscovery of the ``Intermediate 
Population II" first described in the Vatican Proceedings review of O'Connel\ (1958).
The new model is discussed by Gilmore \& Wyse\ (1985) and Wyse \& Gilmore\ (1986). 
Galactic models have been an attractive topic for many research centers, due 
to their importance: Galactic models can be used as a tool to reveal the 
formation and evolution of the Galaxy. For some years there has been a conflict 
among the researchers about the history of our Galaxy. The pioneering work was 
the one of Eggen, Lynden-Bell \& Sandage\ (1962) who argued that the Galaxy 
collapsed in a free-fall time ($\sim 2 \times 10^{8}$ yr). Now, we know that the 
Galaxy collapsed over many Gyr (e.g. Yoshii \& Saio\ 1979; Norris, Bessel \& 
Pickles\ 1985; Norris 1986; Sandage \& Fouts\ 1987; Carney, Latham \& Laird\ 
1990; Norris \& Ryan\ 1991; Beers \& Sommer-Larsen\ 1995) and at least some of 
its components are formed from the merger or accretion of numerous fragments, 
such as dwarf-type galaxies (cf. Searle \& Zinn\ 1978, Freeman \& Bland-Hawthorn\ 
2002, and references therein).           

The researchers use different methods to determine the parameters. Table 1 
summaries the results of these works. One can see that there is an evolution 
for the numerical values of model parameters. The local space density and the 
scaleheight of the thick disc can be given as an example. The evaluations of the 
thick disc have steadily moved toward shorter scaleheights (from 1.45 to 0.65 
kpc, Gilmore \& Reid\ 1983; Chen et al.\ 2001) and higher local densities 
(2-10 per cent). In many studies the range of values for the parameters is large. 
For example, Chen et al.\ (2001) and Siegel et al.\ (2002) give 6.5-13 and 6-10 
per cent, respectively, for the local space density for the thick disc. However, 
one expected the most evolved numerical values for these recent works. That is, 
either the range for this parameter should be small or a single value with a 
small error should be given for it. It seems that they could not choose the most 
appropriate procedures in this topic. In fact, we cited in our previous paper 
(Karaali, Bilir \& Hamzao\u glu\ 2004, hereafter KBH) that the Galactic model 
parameters are mass dependent. Absolute magnitude is reasonable proxy for mass, 
therefore they vary at different absolute magnitude intervals. Hence, the 
parameters cited by the researchers up to recent years which are based on star 
counts cover the range of a series of parameters corresponding to different 
absolute magnitude intervals, therefore either their range  or their errors are 
large. Additionally, as it was cited in our previous paper (KBH), sech$^{2}$ 
density law fits better to the observed density functions for stars with 
absolutely faint magnitudes, $10<M(g^{'})\leq 13$, for the thin disc. We aim to 
use these experiences in the investigation of this field and compare the results 
with those obtained in the field SA 114, almost symmetric relative to the Galactic 
plane. It should be noted that evaluations of photometric parallax (Gilmore \& 
Reid\ 1983, Reid \& Majewski 1993, Siegel et al.\ 2002) have usually broken the 
fits down by absolute magnitude ranges. More importantly, the Besan\c{c}on group 
(e.g. Robin et al.\ 1996) uses very sophisticated models that create multiple 
thin disc populations through population synthesis. This is a much more elegant 
and nuanced way of fitting star count parameters. However, we should mention that 
the method of photometric parallax is, by necessity, a simplified way of 
evaluating star counts.

\section{The density law forms}

Disc structures are usually parameterized in cylindrical coordinates by radial 
and vertical exponentials,

\begin{eqnarray}
\tiny
D_{i}(x,z)=n_{i} exp(-z/H_{i}) exp(-(x-R_{0})/h_{i})
\end{eqnarray}
where $z$ is the distance from Galactic plane, $x$ is the planar distance 
from the Galactic center, $R_{0}$ is the solar distance to the Galactic 
center (8.6 kpc, Buser et al. 1998), $H_{i}$ and $h_{i}$ are the 
scaleheight and scalelength respectively, and $n_{i}$ is the normalized 
local density. The suffix $i$ takes the values 1 and 2, as long as the 
thin and thick discs are considered. It should be noted that the sophisticated 
models of Besan\c{c}on and others use multiple thin discs to account for the range 
of populations (e.g. Robin et al.\ 1996). A similar form uses the sech$^{2}$ 
(or sech) function to parameterize the vertical distribution for the thin disc,

\begin{eqnarray}
D_{1}(x,z)=n_{1}sech^{2}(-z/H^{'}_{1})exp(-(x-R_{o})/h_{1}).
\end{eqnarray}
As the secans hyperbolicus is the sum of two exponentials, $H^{'}_{1}$ is 
not really a scaleheight but has to be compared to $H_{1}$ by dividing it 
with 2: $H_{1}=H^{'}_{1}/2$. We would like to mention that the reason of using 
a sech$^{2}$ law is due to theoretic analysis which indicate that the density 
laws should follow a sech$^{2}$ law for an isothermal sheet. 

The density law for the spheroid component is parameterized in different forms. 
The most common is the de Vaucouleurs\ (1948) spheroid used to describe the 
surface brightness profile of elliptical galaxies. This law has been 
deprojected into three dimensions by Young\ (1976) as 

\begin{eqnarray}
D_{s}(R)=n_{s}~exp[-7.669(R/R_{e})^{1/4}]/(R/R_{e})^{7/8},
\end{eqnarray}
where $R$ is the (uncorrected) Galactocentric distance in spherical 
coordinates, $R_{e}$ is the effective radius and $n_{s}$ is the normalized 
local density. $R$ has to be corrected for the axial ratio $\kappa = c/a$, 

\begin{eqnarray}
R = [x^{2}+(z/\kappa)^2]^{1/2},
\end{eqnarray}
where,
\begin{eqnarray}
z = r \sin b,
\end{eqnarray}
\begin{eqnarray}
x = [R_{o}^{2}+r^{2}\cos^{2} b-2R_{0}r\cos b \cos l]^{1/2}, 
\end{eqnarray}
$r$ being the distance along the line of sight and, $b$ and $l$ the Galactic 
latitude and longitude respectively, for the field under investigation.
The form used by the Basle group is independent of effective radius but 
is dependent on the distance from the Sun to the Galactic centre: 

\begin{eqnarray}
D_{s}(R)=n_{s}~exp[10.093(1-R/R_{o})^{1/4}]/(R/R_{o})^{7/8};
\end{eqnarray}
and alternative formulation is the power law,
\begin{eqnarray}
D_{s}(R)=n_{s}/(a_{o}^{n}+R^{n})
\end{eqnarray}
where $a_{o}$ is the core radius.
	         
Equations (1) and (2) can be replaced by eqs (9) and (10) respectively, 
as long as the vertical direction is considered, where
 
\begin{eqnarray}
D_{i}(z)=n_{i}exp(-z/H_{i}),
\end{eqnarray}
\begin{eqnarray}
D_{1}(z)=n_{1}sech^{2}(-z/H_{1}^{'}).
\end{eqnarray}

\section{The procedure used in this work}

In this work, we used the same procedure cited in our previous paper (KBH), 
i.e. we compared the derived and theoretical space densities per absolute 
magnitude interval, in the vertical direction of the Galactic plane for a 
large absolute magnitude interval $4<M(g^{'})\leq13$, down to the limiting 
magnitude $g^{'}_{0}=20.5$: (i) we separated the stars into different 
populations by their spatial position, as a function of both absolute and 
apparent magnitude; (ii) we tried the exponential and sech$^{2}$ laws for 
comparison of the derived and theoretical space densities for the thin disc 
and we found that a sech$^{2}$ law worked better at magnitudes $10<M(g^{'})\leq13$ 
whereas an exponential density law favors at magnitudes $4<M(g^{'})\leq10$. 
This was also the case in our previous paper (KBH); (iii) we derived model 
parameters for each population individually and for each absolute magnitude 
interval we observed their differences; and (iv) the model parameters were 
estimated by comparison of the derived vertical space densities with the 
combined density laws (eqs. 7 and 9) for stars of all populations. In the 
last process, we obtained two sets of parameters: one for the absolute 
magnitude interval $4<M(g^{'})\leq10$ and the other $4<M(g^{'})\leq13$. 
As we argued in our previous paper, the different behavior of the faint 
stars may produce different values and large ranges for parameters derived 
in starcount studies.

\section{The data and reductions}

\subsection{Observations}
The ELAIS field ($\alpha=16^{h}10^{m}00^{s}, \delta= +54^{o}30^{'}00{''}$; 
$l=84^{o}.27$, $b=+44^{o}.90$; 6.571 deg$^{2}$; epoch 2000) was measured by 
the Isaac Newton Telescope (INT) Wide Field Camera (WFC) mounted at the prime 
focus ($f/3$) of the 2.5-m INT on La Palma, Canary Islands, during seven 
observing runs, namely 1999 April 17; 1999 June 7, 9; 1999 July 16-22; 1999 
August 1-3, 7, 10, 13, 16-17; 1999 September 7-9; 1999 October 3-5, and 2000 
June 24-25. The WFC consists of 4 EVV42 CCDs, each containing 2k $\times$ 4k 
pixels. They are fitted in a L-shaped pattern which makes the camera have 6k 
$\times$ 6k pixels, minus a corner of 2k $\times$ 2k pixels. The WFC has 13.5$\mu$ 
pixels corresponding to 0.33 arcsec pixel$^{-1}$ at the INT prime focus, and 
each covers an area of 22.8 $\times$ 11.4 arcmin$^{2}$ on the sky. This field 
contains 54 sub-fields and each sub-field covers 4 CCDs with a total area of 
0.29 deg$^{2}$. Therefore, the total area of each telescope pointing is 54 $\times$ 
0.29 deg$^{2}$ minus the overlapping area. In our work, the data of only 33 
sub-fields could be used. Hence, the area of the field investigated is 33 
$\times$ 0.29 deg$^{2}$ minus the overlapping area = 6.571 deg$^{2}$. With a 
typical seeing of 1.0-1.3 arcsec on the INT, point objects are well sampled, 
which allows accurate photometry. 

Observations were taken in five bands ($u^{'}_{RGO}$, $g^{'}$, $r^{'}$, $i^{'}$, 
$z^{'}_{RGO}$, where RGO denotes the Royal Greenwich Observatory) with a single 
exposure of 600 s to nominal 5$\sigma$ limiting magnitudes of 23, 25, 24, 23, and 
22 respectively (McMahon et al.\ 2001). However, the limiting magnitudes are 
brighter when stars are considered only. In our work, we determined the $g^{'}_{0}$ 
limiting magnitude for stars by estimating from the star count roll over in Fig. 2 
as $g^{'}_{0}=20.5$. Magnitudes are put on a standard scale using observations of 
ELAIS standard system corresponding to Landolt system \footnote
{http://www.ast.cam.ac.uk/$\sim$wfcsur/technical/photom/}, taken on the 
same night. The accuracy of the preliminary photometric calibrations is $\pm$0.1 
mag. The CCD observations are reduced to the magnitudes by the INT WAS group. 
Three sets of two-colour diagrams, i.e. $(u^{'}-g^{'}, g^{'}-r^{'})$, 
$(g^{'}-r^{'}, r^{'}-i^{'})$ and $(r^{'}-i^{'}, i^{'}-z^{'})$, for 21 sub-fields 
show considerable deviations due to bad reduction hence, we left them out of the 
program. The following processes have been applied to the data for the remaining 
33 sub-fields to obtain a sample of stars with new data available for a model 
parameterization.

\subsection{The overlapping sources, de-reddening of the magnitudes, bright stars, 
and extra-galactic objects}

The data of ELAIS field are provided from the Cambridge Astronomical Survey Unit 
(CASU) \footnote{http://www.ast.cam.ac.uk/$\sim$ wfcsur/release/elaiswfs/}. 
In total, there are 17041 sources in 33 sub-fields in the ELAIS field. It turned 
out that 3027 of these sources are overlapped, i.e. their angular distances are 
less than 1 arcsec to any other source. We omitted them, and so the sample reduced 
to 14014. The $E(B-V)$ colour excess for the sample sources are evaluated by the 
procedure of Schlegel, Finkbeiner \& Davis\ (1998).
\begin{figure}
\center
\resizebox{8cm}{5.5cm}{\includegraphics*{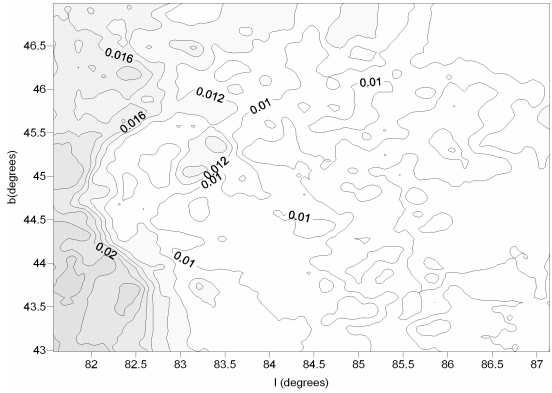}}
\caption[] {$E(B-V)$ colour-excess contours for the field ELAIS as a function 
of Galactic latitude and longitude.}
\end {figure}

\begin{figure}
\center
\resizebox{8cm}{5cm}{\includegraphics*{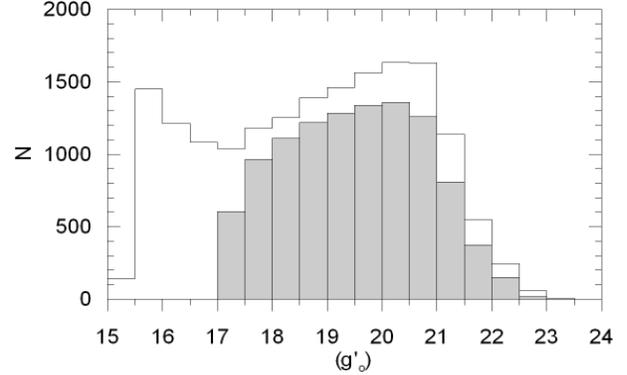}}
\caption[] {An apparent magnitude histogram for all sources (white colour) and 
for only the star sample (black colour).}
\end{figure}

\begin{figure}
\center
\resizebox{6cm}{11.25cm}{\includegraphics*{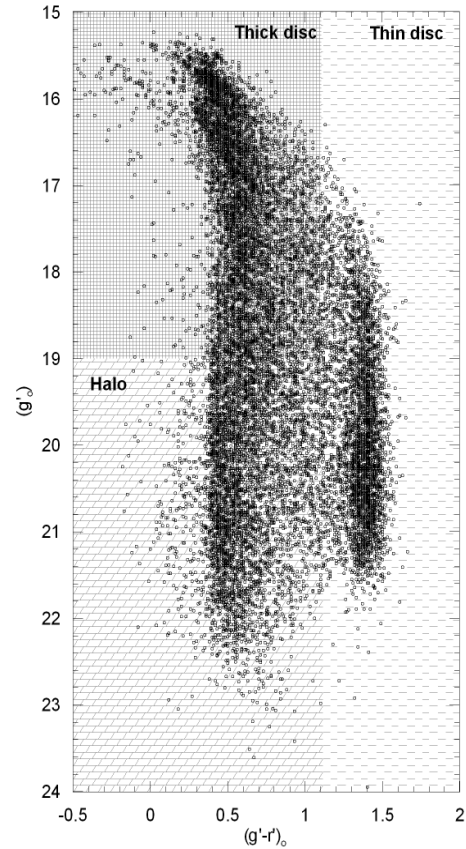}}
\caption[] {Colour-apparent magnitude diagram for the original sample. The shaded 
areas for the regions that correspond to each stellar population is indicated.}
\end{figure}

\begin{figure}
\center
\resizebox{5.6cm}{12.2cm}{\includegraphics*{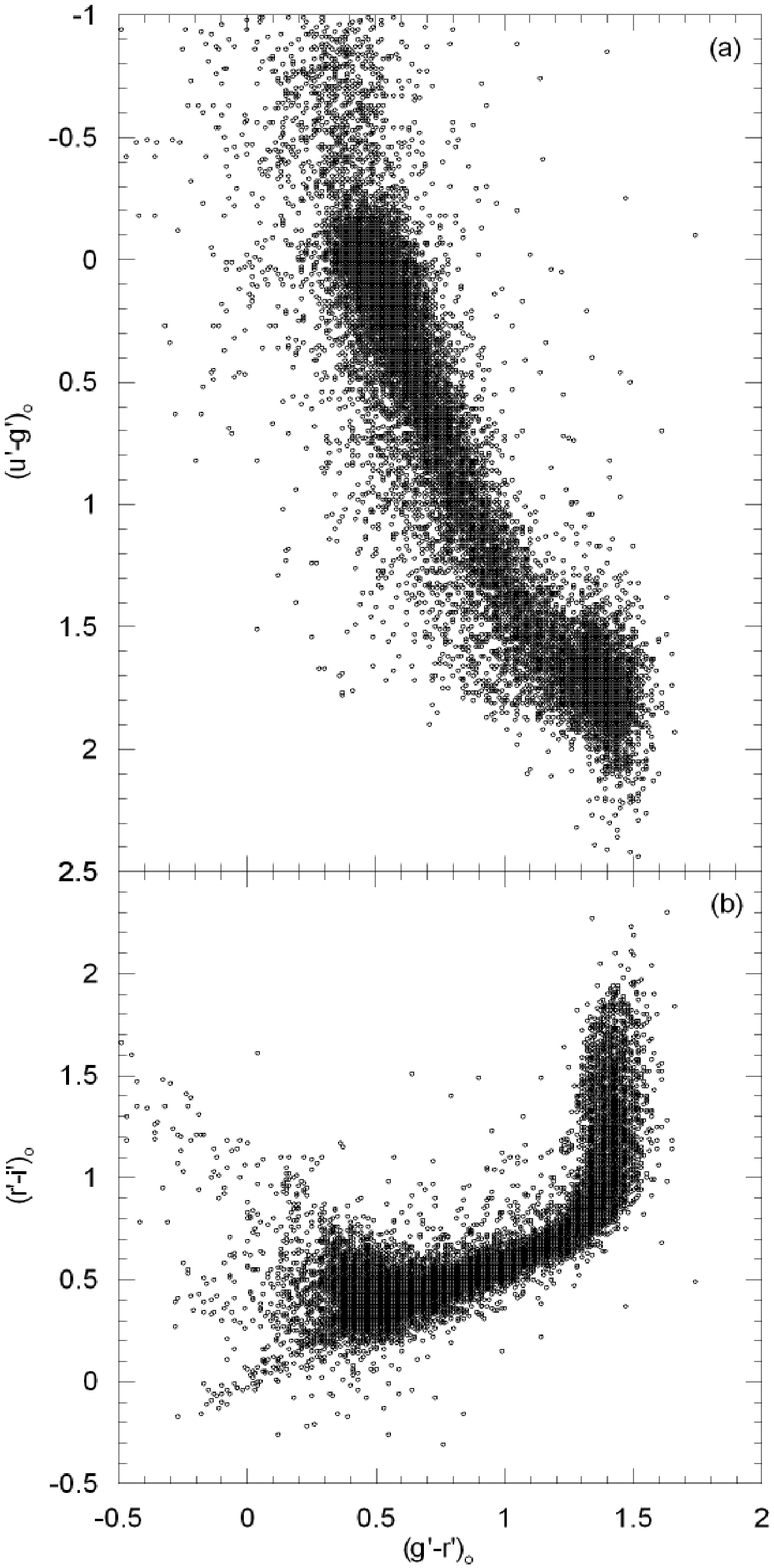}}
\caption[] {Two-colour diagrams for \it sources  \rm with apparent magnitude 
$17<g^{'}_{0}\leq22$: (a) for $(u^{'}-g^{'})_{0}-(g^{'}-r^{'})_{0}$ and 
(b) for ($g^{'}-r^{'})_{0}-(r^{'}-i^{'})_{0}$.}
\end{figure}

\begin{figure}
\center
\resizebox{5.6cm}{12.2cm}{\includegraphics*{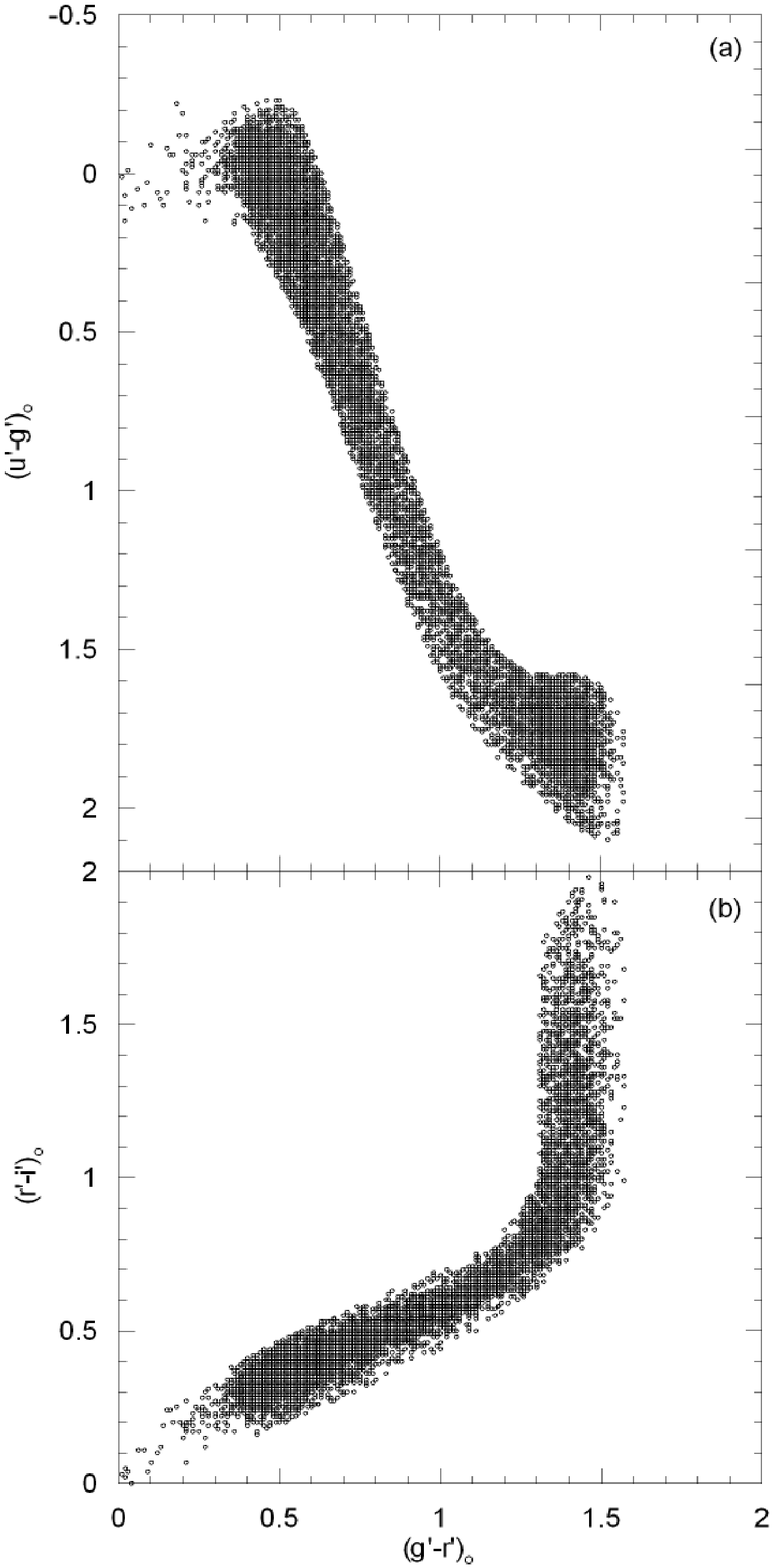}}
\caption[] {Two-colour diagrams for \it stars \rm with apparent magnitude. 
$17<g^{'}_{0}\leq22$: (a) for $(u^{'}-g^{'})_{0}-(g^{'}-r^{'})_{0}$ and 
(b) for ($g^{'}-r^{'})_{0}-(r^{'}-i^{'})_{0}$.}
\end{figure}

\begin{figure}
\center
\resizebox{5.2cm}{10cm}{\includegraphics*{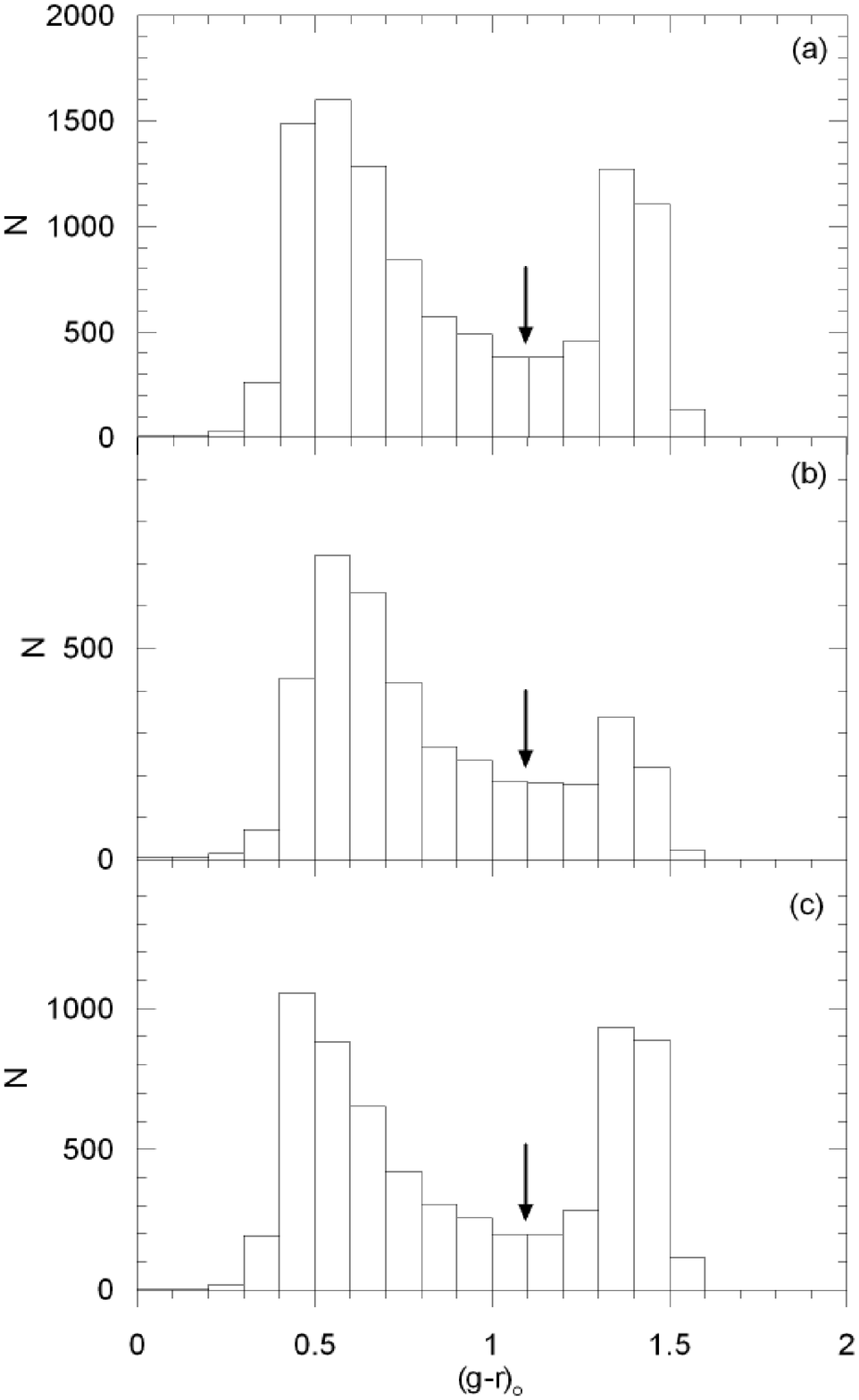}}
\caption[] {A $(g^{'}-r^{'})_{0}$ colour histogram as a function of apparent 
magnitude, for the star sample: (a) for $17<g^{'}_{0}\leq22$, (b) for 
$17<g^{'}_{0}\leq19$ and (c) for $19<g^{'}_{0}\leq22$. The vertical 
downward arrow shows the limit value $(g^{'}-r^{'})_{0}=1.1$ mag which 
separates the thin disc and the thick disc-halo couple.}
\end{figure}
The $E(B-V)$ colour-excess contours for the field are given in Fig. 1 as a 
function of Galactic latitude and longitude. Then, the total absorption 
$A_{V}$ is evaluated by means of the well known equation

\begin{eqnarray}
R_{V}=\frac {A_{V}} {E(B-V)} = 3.1.					
\end{eqnarray}

For Vega bands we used the $R_{\lambda}/R_{V}$ data of Cox\ (2000) for the 
interpolation, where $\lambda$ = 3581, 4846, 6240, 7743, and 8763$\AA$, and 
derived $R_{\lambda}$ from their combination of this with $A_{V}$ (see Table 2, 
KBH). Finally, the dereddened $u_{0}^{'}$, $g_{0}^{'}$, $r_{0}^{'}$, $i_{0}^{'}$, 
and $z_{0}^{'}$ magnitudes were obtained from the original magnitudes and the 
corresponding $R_{\lambda}$.

The histogram for the dereddened apparent magnitude $g_{0}^{'}$ (Fig. 2) and 
the colour-apparent magnitude diagram (Fig. 3) shows that there is a large number 
of saturated sources in our sample. Hence, we excluded sources brighter than 
$g_{0}^{'}=17$ (this bright limit of apparent magnitude matches with the one 
claimed in our previous paper, KBH). However, the two-colour diagrams 
$(u^{'}-g^{'})_{0}-(g^{'}-r^{'})_{0}$ and $(g^{'}-r^{'})_{0}-(r^{'}-i^{'})_{0}$ 
in Fig. 4 indicate that there are also some extragalactic objects, where most 
of them lie towards the blue as claimed by Chen et al.\ (2001) and Siegel et al.\ 
(2002). As claimed in our paper cited above, the star/extragalactic object 
separation based on the ``stellarity parameter'' as returned from the 
SE\tiny XTRACTOR \normalsize routines (Bertin \& Arnouts\ 1996) could not be 
sufficient. We adopted the simulation of Fan\ (1999) in addition to the work cited 
above, to remove the extragalactic objects in our field. Thus we rejected the sources 
with $(u^{'}-g^{'})_{0}<-0.10$ and those which lie outside of the band concentrated 
by most of the sources. After the last process, the number of 6.2 per cent sources 
in the sample-stars-reduced to 10492. The two-colour diagrams 
$(u^{'}-g^{'})_{0}-(g^{'}-r^{'})_{0}$ and $(g^{'}-r^{'})_{0}-(r^{'}-i^{'})_{0}$ for 
the final sample are given in Fig. 5. A few dozen stars with $(u^{'}-g^{'})_{0} 
\sim -0.10$ and $(g^{'}-r^{'})_{0} \sim 0.20$ are probably stars of spectral type A. 

\begin{figure*}
\center
\resizebox{16cm}{8.34cm}{\includegraphics*{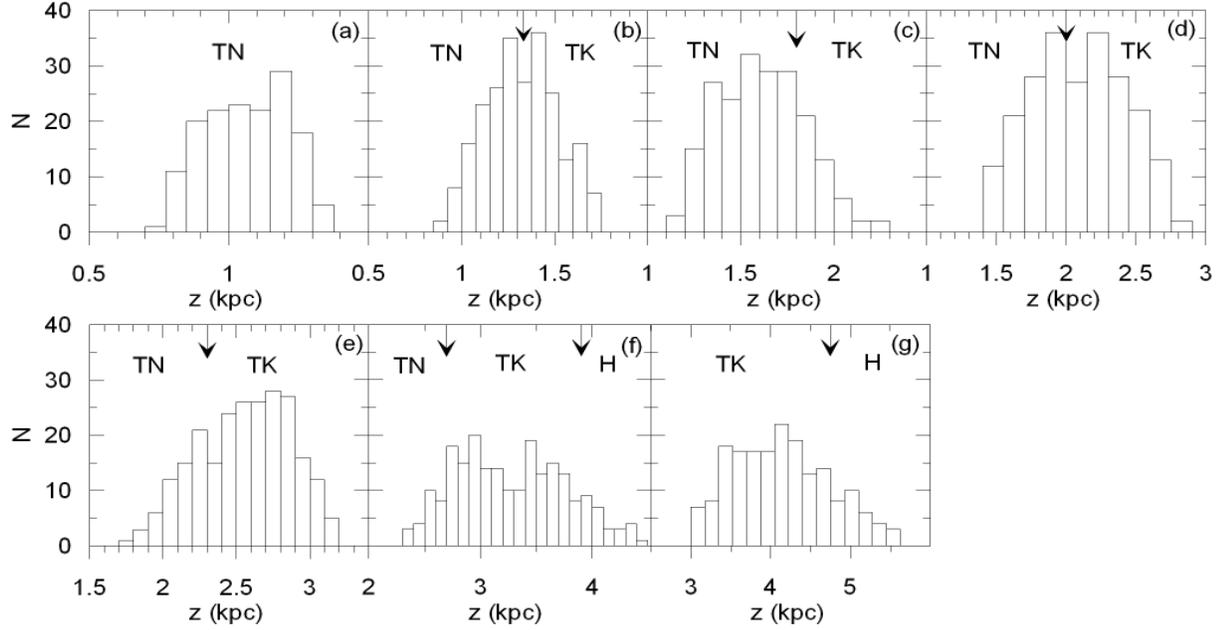}}
\caption[] {Spatial location for stars with absolute magnitude 
$6<M(g^{'})\leq7$ as a function of apparent magnitude: (a)(17.0,17.5], 
(b) (17.5,18.0], (c)(18.0,18.5], (d)(18.5,19.0], (e)(19.0,19.5], 
(f)(19.5,20.0] and (g)(20.0,20.5]. The arrows correspond to the distances from 
the Galactic plane separating the populations (TN: thin disc, TK: thick disk 
and H: halo).}
\end{figure*}

\begin{figure}
\center
\resizebox{6.46cm}{14.5cm}{\includegraphics*{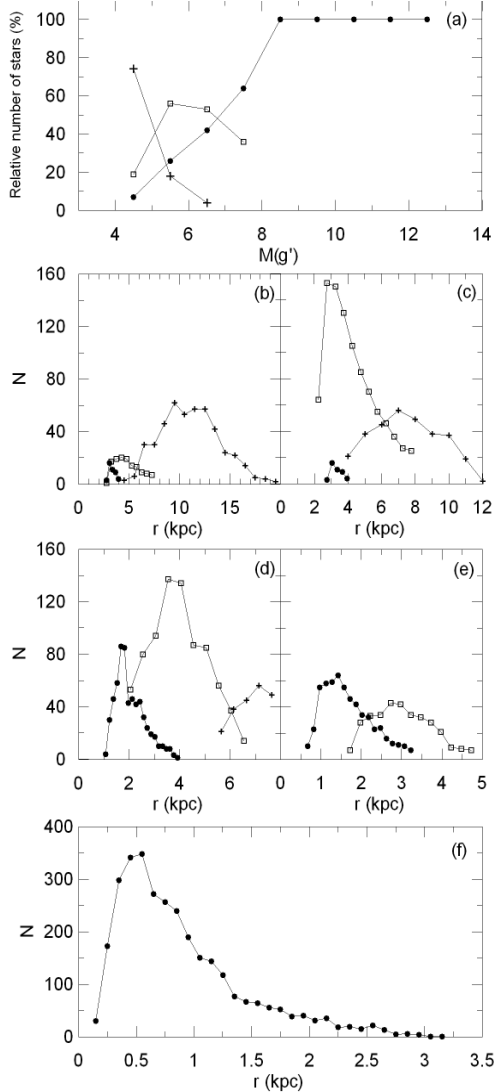}}
\caption[] {Absolute magnitude ranges dominated by different populations 
(panel a) and the break of these contributions down by distance bins for the 
absolute magnitudes $4<M(g^{'})\leq5$, $5<M(g^{'})\leq6$, $6<M(g^{'})\leq7$, 
$7<M(g^{'})\leq8$ and $8<M(g^{'})\leq13$ in panels (b), (c), (d), (e) and (f) 
respectively. Symbols: a plus denotes a halo, a square denotes a thick disc 
and a filled circle denotes a thin disc.}
\end{figure}

\begin{table}
\center
\caption{Dominant regions for three populations: the thin disc, the thick disc 
and the halo, as a function of absolute and apparent magnitudes. The symbol 
$z$ is the distance to the Galactic plane in kpc.}
{\scriptsize
\begin{tabular}{ccccc}
\hline
{$M(g^{'})$}&{$g_{o}^{'}$}&{Thin disc}&{Thick disc}&{Halo}\\
\hline
(12,13] & (17.0,20.5]&   $z\leq0.34$ &$-$              & $-$ \\
(11,12] & (17.0,20.5]&   $z\leq0.55$ &$-$              & $-$ \\
(10,11] & (17.0,20.5]&   $z\leq0.88$ &$-$              & $-$ \\
(9,10]	& (17.0,20.5]&   $z\leq1.40$ &$-$              & $-$ \\  
(8,9]	& (17.0,20.5]&   $z\leq2.24$ &$-$              & $-$ \\
(7,8]	& (17.0,18.0]&   $z\leq1.09$ &$-$              & $-$ \\
	& (18.0,18.5]&   $z\leq1.27$ &$z>1.27$         & $-$ \\
	& (18.5,19.0]&   $z\leq1.34$ &$z>1.34$         & $-$ \\
	& (19.0,19.5]&   $z\leq1.60$ &$z>1.60$         & $-$ \\
	& (19.5,20.0]&   $z\leq1.95$ &$z>1.95$         & $-$ \\
	& (20.0,20.5]&   $z\leq2.32$ &$z>2.32$         & $-$ \\
(6,7]   & (17.0,17.5]&   $z\leq1.40$ &$-$              & $-$ \\
	& (17.5,18.0]&   $z\leq1.38$ &$z>1.38$         & $-$ \\
	& (18.0,18.5]&   $z\leq1.80$ &$z>1.80$         & $-$ \\
	& (18.5,19.0]&   $z\leq1.95$ &$z>1.95$         & $-$ \\
	& (19.0,19.5]&   $z\leq2.30$ &$z>2.30$         & $-$ \\
	& (19.5,20.0]&   $z\leq2.70$ &$2.70<z\leq3.90$ & $z>3.90$\\
	& (20.0,20.5]&   $-$         &$z\leq4.70$      & $z>4.70$\\
(5,6]   & (17.0,17.5]&   $z\leq1.55$ &$z>1.55$         & $-$ \\
        & (17.5,18.0]&   $z\leq1.80$ &$1.80<z\leq2.50$ & $z>2.50$\\
	& (18.0,18.5]&   $z\leq2.20$ &$2.20<z\leq2.90$ & $z>2.90$\\
	& (18.5,19.0]&   $z\leq2.64$ &$2.64<z\leq3.55$ & $z>3.55$\\
	& (19.0,19.5]&   $-$         &$z\leq4.25$      & $z>4.25$\\
	& (19.5,20.0]&   $-$         &$z\leq5.00$      & $z>5.00$\\
	& (20.0,20.5]&   $-$         &$z\leq5.25$      & $z>5.25$\\
(4,5]   & (17.0,17.5]&   $z\leq2.20$ &$z>2.20$         & $-$\\
        & (17.5,18.0]&   $z\leq2.65$ &$2.65<z\leq3.10$ & $z>3.10$\\
	& (18.0,18.5]&   $z\leq3.00$ &$3.00<z\leq4.20$ & $z>4.20$\\
	& (18.5,19.0]&   $-$         &$z\leq4.43$      & $z>4.43$\\
        & (19.0,19.5]&   $-$         &$z\leq5.20$      & $z>5.20$\\
	& (19.5,20.5]&   $-$         &$-$              & $z>5.00$\\
\hline
\end{tabular}
}
\end{table}

\subsection{Absolute magnitudes, distances, population types and density functions}

In the Sloan Digital Sky Survey ($SDSS$) photometry, the blue stars in the range 
$15<g^{*}<18$ are dominated by thick-disc stars with a turn-off $(g^{*}-r^{*})\sim0.33$, 
and for $g^{*}>18$, the Galactic halo, which has a turn-off colour $(g^{*}-r^{*})\sim0.2$, 
becomes significant. Red stars $(g^{*}-r^{*})\geq1.3$, are dominated by thin-disc 
stars for all apparent magnitudes (Chen et al.\ 2001). We used the same procedure 
to demonstrate the three populations (Fig. 3) and to determine the absolute 
magnitudes for stars in each population by appropriate colour-magnitude 
diagrams. In our case, the apparent magnitude which separates the thick disc and 
halo stars seems to be a bit fainter relative to the $SDSS$ photometry, i.e. 
$g^{'}_{0}\sim 19$, and the colour separating the red and bluer stars is slightly 
more blue, i.e. $(g^{'}-r^{'})_{0}=1.1$ (Fig. 6). The absolute magnitudes of 
thick disc and halo stars are evaluated by means of the colour-magnitude diagrams 
of 47 Tuc ([Fe/H]=-0.65 dex, Hesser et al.\ 1987) and M13 ([Fe/H]=-1.40 dex, 
Richer \& Fahlman\ 1986) respectively, whereas for thin disc stars we used the 
colour-magnitude diagram of Lang\ (1992) for population I stars. The colours and 
absolute magnitudes in the $UBV$ system were converted to ELAIS photometry as 
follows: We used two equations for colour transformations from the WEB page of 
CASU\footnote{http://www.ast.cam.ac.uk/$\sim$wfcsur/technical/photom/colours/}:       
   
\begin{eqnarray}
g^{'}-r^{'}=0.908(B-V)+0.048,
\end{eqnarray}
\begin{eqnarray}
g^{'}-B=-0.531(B-V)+0.053.					
\end{eqnarray}
The first equation transforms $B-V$ to $g^{'}-r^{'}$ colour. The second equation can be 
written in the following form which provides $M(g^{'})$ absolute magnitudes: 
\begin{eqnarray}
M(g^{'})=M(B)-0.531(B-V)+0.053.					
\end{eqnarray}
The $M(B)$ absolute magnitudes in eq. 14 were evaluated either by eq. 15 (for the data 
of Lang, 1992) or by eq. 16 (for the data of clusters 47 Tuc and M13):
\begin{eqnarray}
M(B)=M(V)+(B-V),
\end{eqnarray}
\begin{eqnarray}					
M(B)=B-(V-M(V))_{0},					
\end{eqnarray}
where $(V-M(V))_{0}$ is the distance modules of the cluster in question. The distance 
to a star relative to the Sun is carried out by the following formula:
\begin{eqnarray}
[g^{'}-M(g^{'})]_{o} =  5 \log r - 5.					
\end{eqnarray}
The vertical distance to the Galactic plane ($z$) of a star could be evaluated by 
its distance $r$ and its Galactic latitude ($b$) which could be provided by its 
right ascension and declination. 

The precise separation of stars into different populations has been carried out 
by their spatial positions as a function of their absolute and apparent 
magnitudes. The procedure is based on the histograms for distance $z$ from the 
Galactic plane. Fig. 7 gives the histograms for stars with $6<M(g^{'})\leq7$ 
for the apparent $g^{'}_{0}$ magnitude intervals $17<g^{'}_{0}\leq17.5$, 
$17.5<g^{'}_{0}\leq18$, $18<g^{'}_{0}\leq18.5$, $18.5<g^{'}_{0}\leq19$, 
$19<g^{'}_{0}\leq19.5$, $19.5<g^{'}_{0}\leq20$ and $20<g^{'}_{0}\leq20.5$ as an 
example. The vertical arrows show the position that the number of stars decline. 
The distance $z$ corresponding these positions are adopted as the borders of 
three populations of the Galaxy, i.e. they limit the efficiency regions of the 
populations. 

This topic can be clarified in more detail as follows: The dominance regions 
of thin disc, thick disc and halo are at short, intermediate, and large $z$ 
distances respectively. The number counts for thin disc decreases with increasing 
$z$ distance, whereas for thick disc the number counts are very small at short $z$ 
distances, then they increase at intermediate $z$ distances. The decreasing or 
increasing rate is not the same for the two populations. The same case is valid for 
thick disc and halo at large $z$ distances. Hence, statistically, there should be 
drops in the number counts at population transitions. This technique first used in 
the paper of Karaali\ (1994). He showed that the $z$-histograms for three 
populations show a multimodal distribution where the modes at short, intermediate 
and large $z$ distances correspond to thin disc, thick disc and halo stars 
respectively. As shown by Karaali, the agreement of the kinematical distribution of 
the sample stars with their spatial location is a strong confirmation of the 
technique in question.

The technique improved in the recent years (cf. KBH) by introducing the apparent 
magnitude of stars used in the histograms. A population breaks at higher 
$z$-distances when one goes to faint apparent magnitudes, and there are histograms 
where the statistical fluctuations are rather small relative to the deeps at the 
population transitions. These two arguments are the clues in the separation of stars 
into different population types, i.e. thin and thick discs and halo. However, any 
wrong identification of a genuine drop reflects in the value of the parameter in 
question and its corresponding error.

\begin{table*}
\center
\caption{The logarithmic space density function $D^{*}=\log D+10$, for different 
absolute magnitude intervals for the thin disc. $r^{*}=[(r^{3}_{1}+r^{3}_{2})/2]
^{1/3}$ is the centroid distance for the volume $\Delta V_{1,2}$, and 
$z^{*}=r^{*}\sin b$, ($b$) being the Galactic latitude of the field center. The 
other symbols are explained in the text (distances in kpc, volumes in pc$^{3}$).}
{\tiny
\begin{tabular}{rrrrcccccccccccccccccr}
\hline
\multicolumn{4} {r} {$M(g^{'}) \rightarrow$} & \multicolumn{2} {c} {(4,5]}  & \multicolumn{2} {c} {(5,6]} &\multicolumn{2} {c} {(6,7]} & \multicolumn{2} {c} {(7,8]} & \multicolumn{2} {c} {(8,9]} & \multicolumn{2} {c} {(9,10]} & \multicolumn{2} {c} {(10,11]} & \multicolumn{2} {c} {(11,12]} & \multicolumn{2} {c} {(12,13]} \\
\hline
$r_{1}-r_{2}$ & $\Delta V_{1,2}$ & r* & z* &  N & D* &  N & D* &  N & D* &  N & D* & N & D* & N & D* & N & D* & N & D* & N & D* \\
\hline
0.10-0.20 & 4.67 (3) &  0.16 & 0.12 &            &            &            &            &            &            &            &            &            &            &            &            &            &            &         11 &       7.37 &         19 &       7.61 \\ \cline{19-20}
0.20-0.30 & 1.27 (4) &  0.26 & 0.18 &            &            &            &            &            &            &            &            &            &            &          2 &       6.20 &         36 &       7.45 &         90 &       7.85 &         49 &       7.59 \\ \cline{17-18}
0.30-0.40 & 2.47 (4) &  0.36 & 0.25 &            &            &            &            &            &            &            &            &            &            &         20 &       6.91 &        129 &       7.72 &        118 &       7.68 &         35 &       7.15 \\ \cline{15-16} \cline{21-22}
0.40-0.60 & 1.01 (5) &  0.52 & 0.37 &            &            &            &            &            &            &            &            &         22 &       6.34 &        126 &       7.09 &        284 &       7.45 &        238 &       7.37 &         13 &       6.11 \\ \cline{13-14} \cline{19-20}
0.60-0.80 & 1.97 (5) &  0.71 & 0.50 &            &            &            &            &            &            &         16 &       5.91 &        132 &       6.83 &         91 &       6.66 &        238 &       7.08 &         71 &       6.56 &            &            \\
0.80-1.00 & 3.26 (5) &  0.91 &  0.64 &            &            &            &            &            &            &         53 &       6.21 &        165 &       6.70 &         90 &       6.44 &        170 &       6.72 &            &            &            &            \\ \cline{11-12} \cline{17-18}
1.00-1.25 & 6.36 (5) &  1.14 &  0.80 &            &            &            &            &         21 &       5.52 &        107 &       6.23 &        184 &       6.46 &        112 &       6.25 &         68 &       6.03 &            &            &            &            \\ \cline{9-10}
1.25-1.50 & 9.49 (5) &  1.37 &  0.98 &            &            &            &            &         73 &       5.89 &         93 &       5.99 &        113 &       6.08 &         81 &       5.93 &            &            &            &            &            &            \\ \cline{15-16}\
1.50-1.75 & 1.32 (6) &  1.64 &   1.15 &            &            &         13 &       4.99 &        130 &       5.99 &         88 &       5.82 &        114 &       5.94 &         33 &       5.40 &            &            &            &            &            &            \\
1.75-2.00 & 1.76 (6) &  1.88 &   1.33 &            &            &         86 &       5.69 &        122 &       5.84 &         68 &       5.59 &        100 &      5.75 &          5 &       4.45 &            &            &            &            &            &            \\ \cline{7-8} \cline{13-14}
2.00-2.50 & 5.09 (6) &  2.28 &   1.61 &            &            &        167 &       5.52 &        138 &       5.43 &         86 &       5.23 &        117 &       5.36 &            &            &            &            &            &            &            &            \\
2.50-3.00 & 7.59 (6) &  2.77 &   1.96 &         11 &       4.16 &         94 &       5.09 &         72 &       4.98 &         48 &       4.80 &         52 &       4.84 &            &            &            &            &            &            &            &            \\
3.00-3.50 & 1.06 (7) & 3.27 &    2.31 &         21 &       4.30 &         51 &       4.68 &         40 &       4.58 &         22 &       4.32 &            &            &            &            &            &            &            &            &            &            \\ \cline{5-6}
3.50-4.00 & 1.41 (7) & 3.77 &    2.66 &          9 &       3.81 &         26 &       4.27 &         20 &       4.15 &            &            &            &            &            &            &            &            &            &            &            &            \\ \cline{11-12}
4.00-4.50 & 1.81 (7) &  4.26 &   3.01 &          4 &       3.34 &            &            &            &            &            &            &            &            &            &            &            &            &            &            &            &            \\
\hline
\multicolumn{4} {r} {Total} &         45 &            &        437 &            &        616 &            &        581 &            &        999 &            &        560 &            &        925 &            &        528 &            &        116 &            \\
\hline
\end{tabular}  
}
\end{table*}

\begin{table*}
\center
\caption{The logarithmic space density function, $D^{*}=\log D+10$, for different 
absolute magnitude intervals for the thick disc. Symbols as in Table 3.}
\begin{tabular}{rrrrrrrcrcrc}
\hline
\multicolumn{4} {r} {$M(g^{'})\rightarrow$} & \multicolumn{2} {c} {(4,5]} & \multicolumn{2} {c} {(5,6]} & \multicolumn{2} {c} {(6,7]} & \multicolumn{2} {c} {(7,8]} \\
\hline
$r_{1}-r_{2}$ & $\Delta V_{1,2}$ &  r* &  z* &   N &         D* &          N &         D* &          N &         D* &          N &         D* \\
\hline
1.0-1.5 &  1.58 (6) &  1.30 & 0.92 &    &            &            &            &            &            &            &            \\\cline{9-10}
1.5-2.0 &  3.09 (6) &  1.78 & 1.26 &    &            &            &            &         16 &       4.71 &         30 &       4.99 \\\cline{7-8}
2.0-2.5 &  5.09 (6) &  2.28 & 1.61 &    &            &         64 &       5.10 &         80 &       5.20 &         70 &       5.14 \\
2.5-3.0 &  7.59 (6) &  2.77 & 1.96 &  1 &       3.12 &        153 &       5.30 &         73 &       4.98 &         69 &       4.96 \\\cline{5-6}
3.0-3.5 &  1.06 (7) &  3.27 & 2.31 & 17 &       4.21 &        150 &       5.15 &        111 &       5.02 &         75 &       4.85 \\
3.5-4.0 &  1.41 (7) &  3.77 & 2.66 & 10 &       3.85 &        130 &       4.96 &        148 &       5.02 &         46 &       4.51 \\\cline{11-12}
4.0-4.5 &  1.81 (7) &  4.26 & 3.01 & 23 &       4.10 &        105 &       4.76 &        103 &       4.76 &         26 &       4.16 \\
4.5-5.0 &  2.26 (7) &  4.76 & 3.36 & 19 &       3.92 &         85 &       4.58 &         89 &       4.60 &         10 &       3.65 \\
5.0-5.5 &  2.76 (7) &  5.26 & 3.71 & 15 &       3.74 &         70 &       4.40 &         79 &       4.46 &            &            \\
5.5-6.0 &  3.31 (7) &  5.76 & 4.07 & 12 &       3.56 &         55 &       4.22 &         46 &       4.14 &            &            \\\cline{9-10}
6.0-6.5 &  3.91 (7) &  6.26 & 4.42 &  9 &       3.36 &         46 &       4.07 &         27 &       3.84 &            &            \\
6.5-7.0 &  4.56 (7) &  6.76 & 4.77 &  7 &       3.19 &         36 &       3.90 &          5 &       3.04 &            &            \\
7.0-8.0 &  1.13 (8) &  7.53 & 5.32 &  9 &       2.90 &         25 &       3.35 &            &            &            &            \\
8.0-9.0 &  1.45 (8) &  8.53 & 6.02 &    &            &         27 &       3.27 &            &            &            &            \\
\hline
\multicolumn{4} {r} {Total} &        122 &            &        946 &            &        777 &            &        326 &           \\
\hline
\end{tabular}  
\end{table*}

Table 2 gives the full set of absolute and apparent magnitude intervals and the 
efficiency regions of the populations. The distance over which a population, the 
thin disc for example, dominates increases with declining absolute magnitude. 
That is, the three populations are not squeezed into small isolated volumes. The 
same holds also when one goes to apparently faint magnitudes in an absolute 
magnitude interval. These findings were cited also in our previous paper (KBH) 
and are consistent with the results of Reid \& Majewski\ (1993), who argued that 
the thick disc extends up to $z\sim4$ kpc, a distance from the Galactic plane 
where halo stars cannot be omitted. Halo stars dominate the absolutely bright 
intervals, thick disc stars indicate the intermediate brightness intervals and 
thin disc stars indicate the faint intervals, as expected (Fig. 8a). If we break 
these contributions down by distance bins, we would reveal that the efficient 
region for each population shifts to shorter distances relative to the Sun, when 
one goes from absolutely bright to absolutely faint magnitudes (Fig. 8b-f). For 
example, thin disc is efficient at $r\sim 1.5$ kpc for the absolutely magnitudes 
$6<M(g^{'})\leq7$ and $7<M(g^{'})\leq8$ whereas the efficiency shifts to 
$r\sim0.5$ kpc for the interval $8<M(g^{'})\leq13$.    

\begin{table}
\center
\caption{The logarithmic space density function, $D^{*}=\log D+10$, for different 
absolute magnitude intervals for the halo. Symbols as in Table 3.}
{\tiny
\begin{tabular}{lrrrrrrcrc}
\hline
\multicolumn{4} {r} {$M(g^{'})\rightarrow$} & \multicolumn{2} {c} {(4,5]} &\multicolumn{2} {c} {(5,6]}& \multicolumn{2} {c} {(6,7]} \\
\hline
$r_{1}-r_{2}$ & $\Delta V_{1,2}$ &  r* &    z* &  N &         D* &     N &         D* &          N &         D* \\
\hline
 3-4   & 1.41 (7) &  3.77 &  2.66 &            &            &         11 &       3.89 &            &            \\ \cline{5-6}
 4-6   & 1.01 (8) &  5.19 &  3.66 &          9 &       2.95 &         75 &       3.87 &         24 &       3.37 \\ \cline{9-10}
 6-8   & 1.97 (8) &  7.14 &  5.04 &         60 &       3.48 &         95 &       3.68 &         41 &       3.32 \\
 8-10  & 3.26 (8) &  9.11 &  6.43 &        108 &       3.52 &         90 &       3.44 &            &            \\
10-15  & 1.58 (9) & 12.98 &  9.16 &        231 &       3.16 &         34 &       2.33 &            &            \\ \cline{7-8}
15-17.5& 1.32 (9) & 16.35 & 11.54 &         41 &       2.49 &            &            &            &            \\
\hline
\multicolumn{4} {r} {Total}       &        449 &            &        305 &            &         65 &            \\
\hline
\end{tabular}  
}  
\end{table}

The logarithmic density functions, $D^{*}=\log D+10$, are given in Tables 3-5 
and Figures 9-11 for different absolute magnitudes for three populations,  
where: $D=N/ \Delta V_{1,2}$; 
$\Delta V_{1,2}=(\pi/180)^{2}(\sq/3)(r_{2}^{3}-r_{1}^{3})$; $\sq$ denotes the size 
of the field (6.571 deg$^{2}$); $r_{1}$ and $r_{2}$ denote the limiting distance 
of the volume $\Delta V_{1,2}$; $N$ denotes the number of stars (per unit 
absolute magnitude); $r^{*}=[(r^{3}_{1}+r^{3}_{2})/2]^{1/3}$ is the centroid 
distance for the volume $\Delta V_{1,2}$; and $z^{*}=r^{*}\sin b$, ($b$) being the 
Galactic latitude of the field center. The horizontal thick lines, in Tables 3-5, 
corresponding to the limiting distance of completeness ($z_{l}$) are evaluated by 
the following equations:

\begin{eqnarray}
[g^{'}-M(g^{'})]_{o} =  5 \log r_{l} - 5, \\
z_{l}=r_{l}\sin b,					
\end{eqnarray}
where $g_{0}^{'}$ is the limiting apparent magnitude (17 and 20.5 for the bright 
and faint stars respectively), $r_{l}$ the limiting distance of completeness relative 
to the Sun and $M(g^{'})$ is the appropriate absolute magnitude $M_{1}$ or $M_{2}$ 
for the absolute magnitude interval $M_{1}-M_{2}$ considered. 

\begin{figure*}
\center
\resizebox{16cm}{9cm}{\includegraphics*{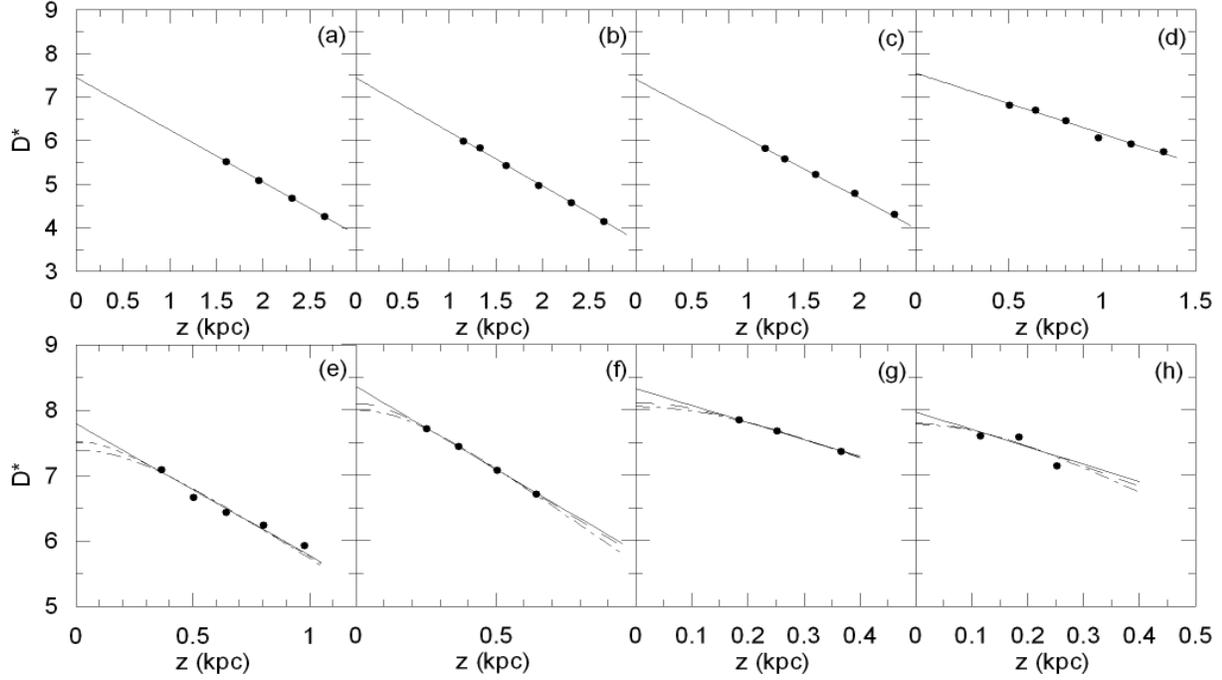}}
\caption[] {Comparison of the observed space density function with the density 
laws for different absolute magnitude intervals for the thin disc. (a) (5,6], 
(b) (6,7], (c) (7,8], (d) (8,9], (e) (9,10], (f) (10,11], (g) (11,12] and
(h) (12,13]. The continuous curve represents the exponential law, the dashed 
curve, represents the $sech$ law and dashed-dot curve represents the sech$^{2}$ 
law.}
\end{figure*}

\begin{figure}
\center
\resizebox{8cm}{8.8cm}{\includegraphics*{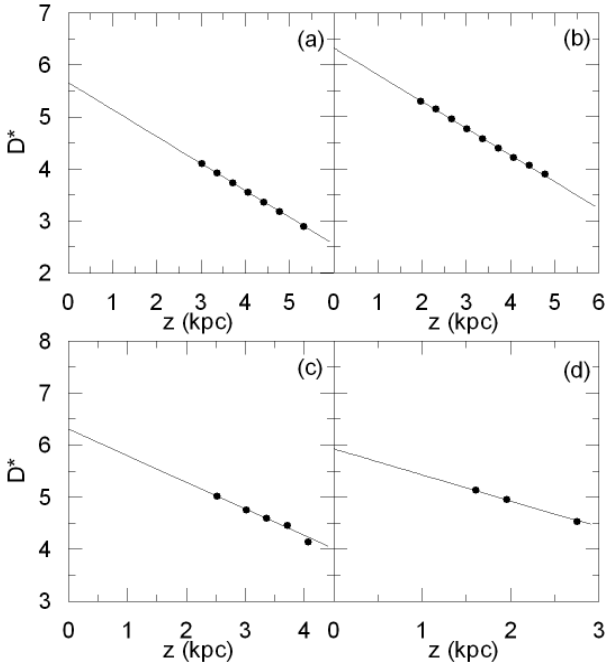}}
\caption[] {Comparison of the derived space density function with the 
exponential density law for different absolute magnitude intervals for 
the thick disc. (a) (4,5], (b) (5,6], (c) (6,7] and (d) (7,8].}
\end{figure}

\begin{figure}
\center
\resizebox{8cm}{8.8cm}{\includegraphics*{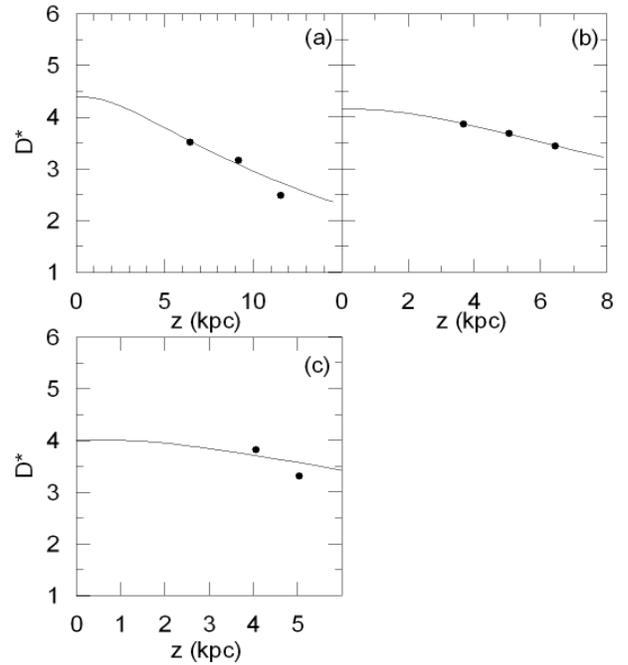}}
\caption[] {Comparison of the derived space density function with the de 
Vaucouleurs density law for different absolute magnitude intervals for the 
halo. (a) (4,5], (b) (5,6] and (c) (6,7].}
\end{figure}

We acknowledge that in this work we have used a simple method. We have postulated 
mono-metallic stellar populations with no abundance gradient, and we have not applied 
any correction for binarism, contamination by compact galaxies or giant/sub-giants 
neither. We are making a simplified evaluation. 

\begin{table*}
\center
\caption{Galactic model parameters for different absolute magnitude 
intervals for the thin disc resulting from the comparison of observed 
logarithmic space densities with a (unique) density law (Fig. 9). The 
columns give: absolute magnitude interval $M(g^{'})$, the density law, 
the logarithmic local space density $n^{*}$, scaleheight for sech or 
sech$^{2}$ density law $H_{1}^{'}$ (in the parenthesis), the scaleheight 
for exponential density law $H$, $\chi^{2}_{min}$, the standard deviation 
$s$ and the local space density for $Hipparcos$ $\odot$.}
{\scriptsize
\begin{tabular}{cccccrc}
\hline
$M(g^{'})$ & Density law & $n^{*}$ &  $(H_{1}^{'})~~H$&  $\chi^{2}_{min}$&$s$& $\odot$ \\
\hline
(12,13] &      exp   & $7.97^{+0.10}_{-0.10}$ &  $163^{+42}_{-28}$     &  7.33  & $\pm$ 0.13 & 8.05\\
        &     sech   & $7.80^{+0.09}_{-0.09}$ &$(227)~~137^{+31}_{-22}$&  6.42  &  0.10 & \\
        & sech$^{2}$ & $7.78^{+0.09}_{-0.09}$ &$(428)~~214^{+45}_{-31}$&  5.70  &  0.11 & \\
(11,12] &      exp   & $8.33^{+0.01}_{-0.01}$ &  $167^{+1}_{-1}$       &  0.09  &  0.01 & 7.92\\
        &     sech   & $8.11^{+0.01}_{-0.01}$ &$(254)~~153^{+1}_{-1}$  &  0.01  &  0.00 & \\
        & sech$^{2}$ & $8.05^{+0.01}_{-0.01}$ &$(514)~~257^{+3}_{-2}$  &  0.07  &  0.01 & \\
(10,11] &      exp   & $8.36^{+0.01}_{-0.01}$ &  $172^{+1}_{-2}$       &  0.58  &  0.01 & 7.78\\
        &     sech   & $8.10^{+0.01}_{-0.01}$ &$(275)~~166^{+1}_{-2}$  &  0.13  &  0.01 & \\
        & sech$^{2}$ & $8.00^{+0.01}_{-0.01}$ &$(590)~~295^{+2}_{-3}$  &  0.55  &  0.01 & \\
 (9,10] &      exp   & $7.79^{+0.07}_{-0.07}$ &   $216^{+14}_{-13}$    &  16.90 &  0.09 & 7.63\\
        &     sech   & $7.52^{+0.07}_{-0.07}$ &$(350)~~211^{+16}_{-13}$&  19.48 &  0.09 & \\
        & sech$^{2}$ & $7.39^{+0.08}_{-0.08}$ &$(770)~~385^{+35}_{-28}$&  25.95 &  0.11 & \\
  (8,9] &      exp   & $7.55^{+0.05}_{-0.05}$ &$313^{+15}_{-13}$       &  12.70 &  0.06 & 7.52\\
  (7,8] &      exp   & $7.41^{+0.02}_{-0.02}$ &$318^{+3}_{-3}$         &   1.59 &  0.04 & 7.48\\
  (6,7] &      exp   & $7.44^{+0.03}_{-0.03}$ &$351^{+6}_{-6}$         &   2.49 &  0.03 & 7.47\\
  (5,6] &      exp   & $7.44^{+0.01}_{-0.01}$ &$363^{+1}_{-1}$         &   0.03 &  0.01 & 7.47\\
\hline
\end{tabular}
}  
\end{table*}

\section{Galactic model parameters}

We estimated Galactic model parameters by comparison of the derived space 
density functions, with the density laws both independently for each population 
as a function of absolute magnitude and simultaneously for all stellar 
populations. 

\subsection{Absolute magnitude dependent Galactic model parameters} 

The thin disc density laws were fitted with the additional constraint of 
producing local densities consistent with those derived from $Hipparcos$    
(Jahreiss \& Wielen\ 1997), a procedure applied in our previous paper (KBH). 
It was discovered that the sech$^{2}$ law fitted better for the intervals  
$10<M(g^{'})\leq11$ and $11<M(g^{'})\leq12$ confirming our results in our paper 
mentioned above, however, contrary to our expectation, the exponential law 
fitted better for the interval $12<M(g^{'})\leq13$ and the sech law fitted 
better for the interval $9<M(g^{'})\leq10$, whereas the exponential law was 
favourite in our previous paper (KBH, Table 6 and Fig. 9). The comparison for 
absolutely bright intervals, i.e. $8<M(g^{'})\leq9$, $7<M(g^{'})\leq8$, 
$6<M(g^{'})\leq7$ and $5<M(g^{'})\leq6$ is carried out with the exponential 
law, as in our previous paper cited above. The scaleheight for thin disc 
increases monotonically from 163 to 363 pc when one goes from the absolute 
magnitude interval $12<M(g^{'})\leq13$ to $5<M(g^{'})\leq6$, with the exception 
scaleheight H=211 pc for the interval $9<M(g^{'})\leq10$ which is less than 
the one for the interval $10<M(g^{'})\leq11$, i.e. H=295 pc. As cited above, 
$9<M(g^{'})\leq10$ is the unique absolute magnitude interval where sech density 
law fitted better with the derived space densities in our work, and it is a 
transition interval between those for which either exponential law (for bright 
intervals) or sech$^{2}$ law (for fainter intervals) fitted better. All 
scaleheights are equivalent to the exponential law scaleheights. The local 
space density for thin disc, for different absolute magnitude intervals, is 
consistent with the $Hipparcos'$ one (Table 6).

For the thick disc, the derived logarithmic space density functions are compared 
with the exponential density law for the absolute magnitude intervals 
$7<M(g^{'})\leq8$, $6<M(g^{'})\leq7$, $5<M(g^{'})\leq6$ and $4<M(g^{'})\leq5$, 
(Table 7 and Fig. 10). The range for the scaleheight is rather small, 839-867 pc 
and the scaleheight itself is flat within the quoted uncertainties. The local space 
density relative to the local space density of thin disc ($n_{2}/n_{1}$) could 
not be given for the interval $4<M(g^{'})\leq5$ due to lack of local space density 
for this interval for the thin disc. For the intervals $5<M(g^{'})\leq6$ and 
$6<M(g^{'})\leq7$, $n_{2}/n_{1}$ is 7.59 and 7.41 per cent respectively, equivalent 
to the updated numerical values, whereas for the faintest interval, $7<M(g^{'})\leq8$, 
$n_{2}/n_{1}$=3.31 per cent is close to the original value (Gilmore \& Wyse\ 1985).
   
The derived logarithmic space density functions for the halo are compared with 
the de Vaucouleurs density law for the absolute magnitude intervals 
$6<M(g^{'})\leq7$, $5<M(g^{'})\leq6$ and $4<M(g^{'})\leq5$ (Table 8 and Fig. 
11). The local space density relative to the thin disc ($n_{3}/n_{1}$) could 
not be given for the interval $4<M(g^{'})\leq5$ due to the reason cited above. 
For the intervals $6<M(g^{'})\leq7$ and $5<M(g^{'})\leq6$, $n_{3}/n_{1}$=0.04 
and 0.06 per cent respectively. The numerical values for the axial ratio 
$\kappa$ for two intervals are close to each other, i.e. $\kappa=0.78$ and 
$\kappa=0.73$ for $6<M(g^{'})\leq7$ and $5<M(g^{'})\leq6$ respectively, but 
a bit less for the interval $4<M(g^{'})\leq5$, $\kappa=0.60$, consistent with 
the previous ones within the uncertainties however. 

The parameters derived for three populations have been tested by the luminosity 
function (Fig. 12), where $\varphi^{*}(M)$ is the total of the local space 
densities for three populations. The local space densities for the thick disc 
and the halo are presented in Table 9. The local space densities of $Hipparcos$ 
were converted to ELAIS colours by the combination of eqs (13) and (15) which 
give the following relation between $M(g^{'})$ and $M(V)$ absolute magnitudes: 
\begin{eqnarray}
M(g^{'})=M(V)+0.469(B-V)+0.053
\end{eqnarray}
There is a good agreement between our luminosity function and that of $Hipparcos$ 
(Jahreiss \& Wielen\ 1997). Also the error bars are rather short. We used the 
procedure of Phleps et al.\ (2000) for the error estimation in Tables 6-8 (above)
and Tables 13 and 14 (in the following sections), i.e. changing the values of 
the parameters until $\chi^{2}_{min}$ increases or decreases by 1.  

\begin{table}
\center
{\scriptsize
\caption{Galactic model parameters for the thick disc. $n_{2}/n_{1}$ indicates the 
local space density for the thick disc relative to the thin disc. Other symbols are 
same as in Table 6.}
\begin{tabular}{cccccc}
\hline
$M(g^{'})$ & $n^{*}$ & $H$ (pc)& $\chi^{2}_{min}$ & $s$ & $n_{2}/n_{1}$ (per cent)\\
\hline
(7,8] &$5.93^{+0.03}_{-0.03}$ &$867^{+24}_{-21}$ &  0.32 & $\pm$ 0.05 & 3.31 \\
(6,7] &$6.31^{+0.03}_{-0.03}$ &$849^{+14}_{-16}$ &  2.88 &       0.05 & 7.41 \\
(5,6] &$6.32^{+0.01}_{-0.01}$ &$845^{+9}_{-7}$   &  0.88 &       0.02 & 7.59 \\
(4,5] &$5.66^{+0.01}_{-0.01}$ &$839^{+2}_{-2}$   &  0.01 &       0.01 &  $-$\\
\hline
\end{tabular}
}  
\end{table}

\begin{table}
\center
\caption{Galactic model parameters for the halo. $\kappa$ and $n_{3}/n_{1}$ 
give the axial ratio and the local space density for the halo relative to the 
thin disc, respectively. Other symbols are as in Table 6.}
{\scriptsize
\begin{tabular}{cccccc}
\hline
$M(g^{'})$& $n^{*}$ & $\kappa$ &$\chi^{2}_{min}$&  $s$ & $n_{3}/n_{1}$ (per cent)\\
\hline
(6,7] &$4.06^{+0.18}_{-0.14}$ & $0.78^{+0.22}_{-0.20}$ & 10.21 & $\pm$ 0.26 & 0.04 \\
(5,6] &$4.19^{+0.01}_{-0.01}$ & $0.73^{+0.02}_{-0.01}$ &  0.14 &       0.01 & 0.06 \\
(4,5] &$4.43^{+0.08}_{-0.09}$ & $0.60^{+0.06}_{-0.05}$ & 21.08 &       0.24 & $-$  \\
\hline
\end{tabular}
}  
\end{table}

\begin{table}
\center
\caption{Local luminosity functions for thick disc ($\varphi^{*}(M)_{TK}$) and halo 
($\varphi^{*}(M)_{H}$). The luminosity function of $Hipparcos$ is also given in 
the last column.}
\begin{tabular}{cccc}
\hline
$M(g^{'})$ & $\varphi^{*}(M)_{TK}$ & $\varphi^{*}(M)_{H}$ & $\odot$\\
\hline
(7,8]  & 5.93 & $-$  & 7.47\\
(6,7]  & 6.31 & 4.06 & 7.47\\
(5,6]  & 6.32 & 4.19 & 7.47\\
(4,5]  & 5.66 & 4.43 & 7.30\\
\hline
\end{tabular}  
\end{table}

\begin{figure}
\center
\resizebox{8cm}{5cm}{\includegraphics*{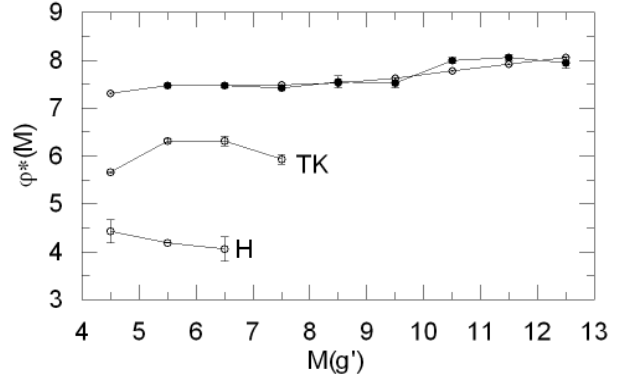}}
\caption[] {The local luminosity function obtained from combining the local 
space densities for the thin and thick discs and the halo, resulting from 
comparison of the derived space density function with the density laws, for 
different absolute magnitude intervals. The ``$\odot$" symbols show the 
$Hipparcos$ values. TK and H corresponds to only thick disc and halo local 
luminosity, respectively.}
\end{figure}

\subsection{Model parameter estimation by simultaneous comparison to the 
Galactic stellar populations}

We estimated the model parameters for three populations simultaneously by 
comparison of the combined derived space density functions with the combined 
density laws. We carried out this work for two sets of absolute magnitude 
intervals, $4<M(g^{'})\leq10$ and $4<M(g^{'})\leq13$. The fit for the second 
interval was done due to our experience that model parameters are absolute 
magnitude dependent and that it covers the thin-disc stars with 
$10<M(g^{'})\leq13$, the density functions of which behave differently from 
the density functions for stars with other absolute magnitudes. Actually, we 
will see in the following that the luminosity function for the absolute 
magnitude interval $4<M(g^{'})\leq13$ differs from the one for stars with 
$4<M(g^{'})\leq10$ considerably (Fig. 14 and Fig. 16). The number of stars as 
a function of distance $r$ relative to the Sun for nine absolute magnitude 
intervals are given in Table 10 and the density functions per unit absolute 
magnitude interval evaluated by these data are shown in Tables 11 and 12 for 
the intervals $4<M(g^{'})\leq10$ and $4<M(g^{'})\leq13$ respectively.

\begin{table*}
\center
\caption{Number of stars as a function of distance $r$ relative to the Sun 
for nine absolute magnitude intervals (distances in kpc). Horizontal thick 
lines correspond the limiting distance of completeness.}
\begin{tabular}{rrrrrrrrrr}
\hline
$M(g^{'}) \rightarrow$ &       (4,5] &      (5,6] &      (6,7] &      (7,8] &      (8,9] &     (9,10] &    (10,11] & (11,12] & (12,13] \\
$r_{1}-r_{2}$&        N &          N &          N &          N &          N &          N &          N &          N & N \\
\hline
   0.0-0.2 &            &            &            &            &            &            &            &         11 & 19 \\ \cline{9-10}
   0.2-0.4 &            &            &            &            &            &         22 &        165 &        208 & 84 \\ \cline{7-8} \cline{10-10}
   0.4-0.7 &            &            &            &          4 &         67 &        177 &        412 &        289 & 13 \\ \cline{6-6} \cline{9-9}
   0.7-1.0 &            &            &            &         65 &        252 &        130 &        280 &         20 &    \\ \cline{5-5} \cline{8-8}
   1.0-1.5 &            &            &         94 &        200 &        297 &        193 &         68 &            &    \\ \cline{4-4} \cline{7-7}
   1.5-2.0 &            &         99 &        268 &        186 &        214 &         38 &            &            &    \\ \cline{3-3} 
   2.0-2.5 &            &        231 &        218 &        156 &        117 &            &            &            &    \\ \cline{6-6}
   2.5-3.0 &         12 &        247 &        145 &        117 &         52 &            &            &            &    \\ \cline{2-2}
   3.0-4.0 &         55 &        368 &        319 &        143 &            &            &            &            &    \\ \cline{5-5}
   4.0-5.0 &         46 &        234 &        192 &         36 &            &            &            &            &    \\ \cline{4-4}
   5.0-7.5 &        102 &        319 &        215 &            &            &            &            &            &    \\ 
   7.5-10.0&        124 &        156 &          7 &            &            &            &            &            &    \\ \cline{3-3}
  10.0-12.5&        146 &         34 &            &            &            &            &            &            &    \\
  12.5-15.0&         85 &            &            &            &            &            &            &            &    \\\cline{2-2}
  15.0-17.5&         41 &            &            &            &            &            &            &            &    \\
     Total &        611 &       1688 &       1458 &        907 &        999 &        560 &        925 &        528 &116 \\
\hline
\end{tabular}
\end {table*}

\subsubsection {Model parameters by means of absolute magnitudes 
$4<M(g^{'})\leq10$}

The combined derived densities per absolute magnitude interval for three 
populations, the thin and thick discs and the halo for stars with 
$4<M(g^{'})\leq10$ (Table 11), are compared with the combined density laws 
(Fig. 13). The derived parameters are given in Table 13. All these 
parameters are in agreement with the ones given in Table 1, and they lie 
between two corresponding parameters cited in Section 5.1, except the 
scaleheight of thick disc, 760 pc, which is rather smaller than the 
scaleheight 839 pc, the smallest one in Table 7. Thus, the scaleheight 269 
pc for the thin disc is between the ones for absolute magnitude intervals 
$10<M(g^{'})\leq11$ and $11<M(g^{'})\leq12$, and the logarithmic local space 
density 7.51 is almost equal to the corresponding one for the absolute 
magnitude interval $9<M(g^{'})\leq10$. For the thick disc, the local space 
density relative to thin disc, 6.46 per cent, is between the local space 
densities for the absolute magnitude intervals $6<M(g^{'})\leq7$ and 
$7<M(g^{'})\leq8$. It is interesting that the local space density for the halo 
relative to thin disc, 0.08 per cent, is almost equal to the one for the 
absolute magnitude interval $5<M(g^{'})\leq6$ in Table 8, however, the axial 
ratio $\kappa=0.55$ is considerably smaller than the ones in the same table. 
The resulting luminosity function (Fig. 14) from the comparison of the model 
with these parameters and the combined derived density functions per absolute 
magnitude interval is in agreement with the one of $Hipparcos$ (Jahreiss 
\& Wielen\ 1997). However, the error bars are longer than the ones in Fig. 12, 
particularly for the faint magnitudes.

\begin{figure}
\center
\resizebox{8cm}{5cm}{\includegraphics*{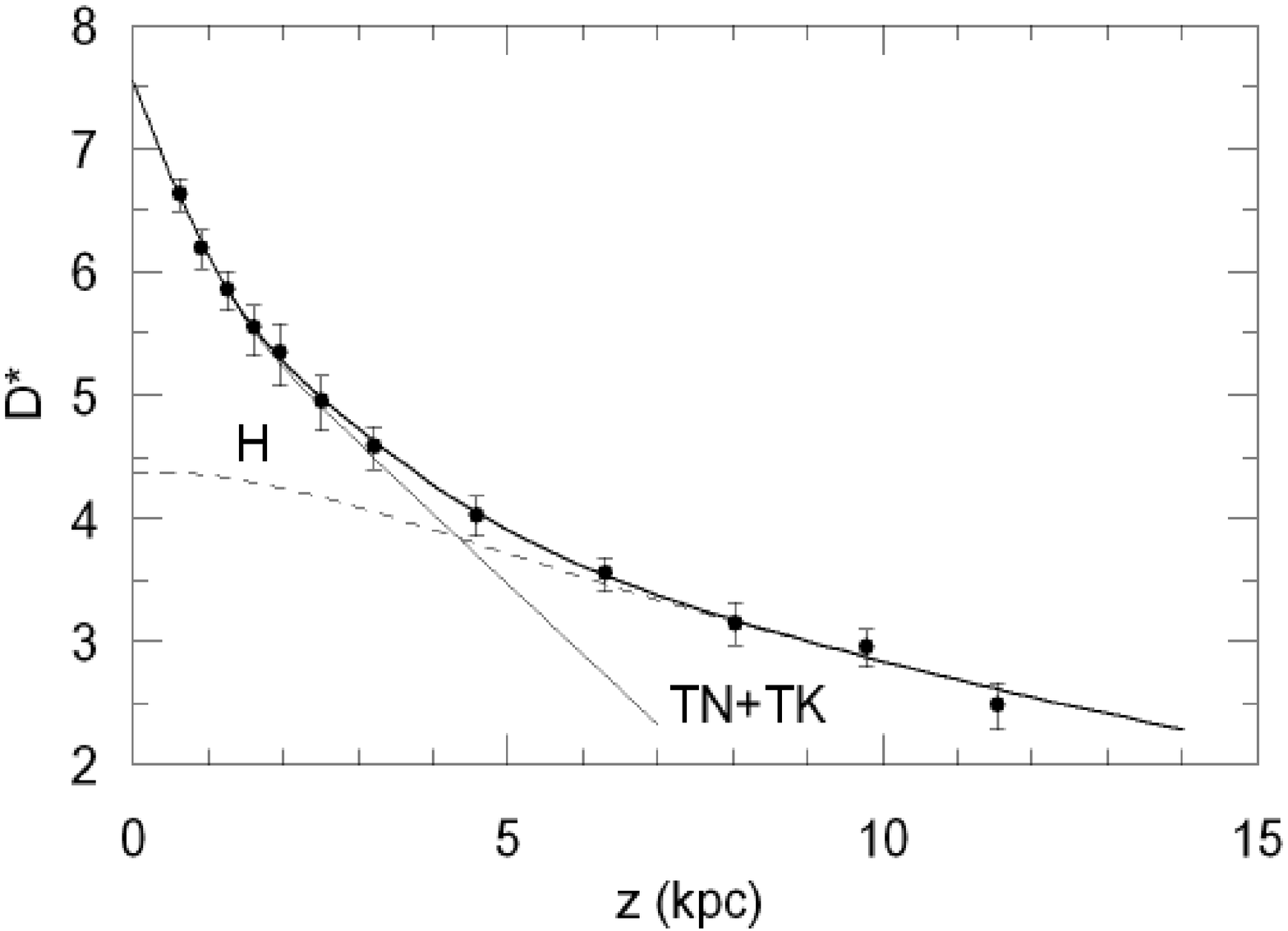}}
\caption[] {Comparison of the derived and combined space density function 
for the thin and thick discs and the halo with the combined density law, for 
stars with $4<M(g^{'})\leq10$.}
\end{figure}

\begin{figure}
\center
\resizebox{8cm}{5.5cm}{\includegraphics*{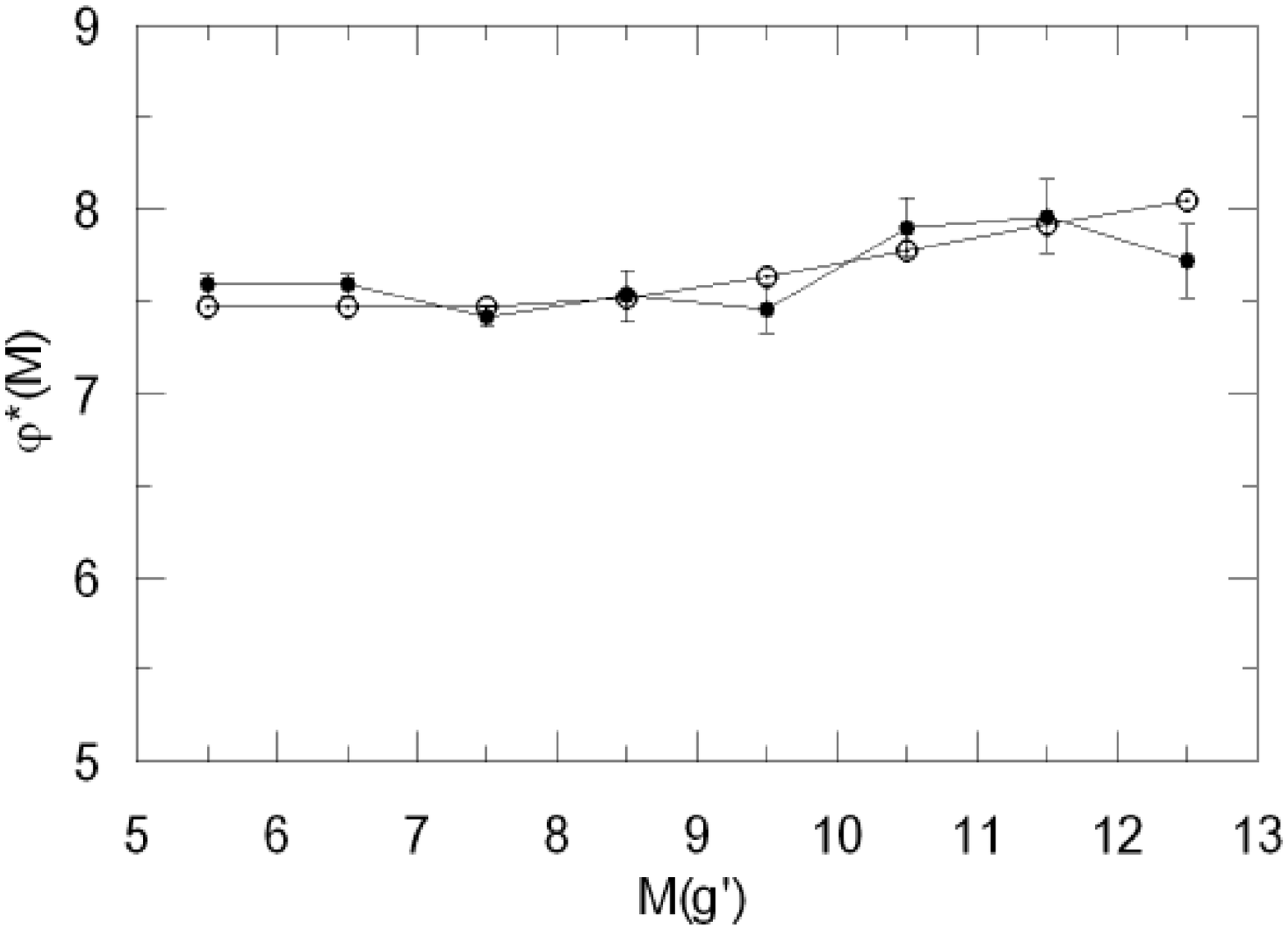}}
\caption[] {The local luminosity function resulting from the comparison of the 
combined derived space density function with the combined density law, for stars with 
$4<M(g^{'})\leq10$.}
\end{figure}

\begin{table}
\center
\caption{The logarithmic space density function, $D^{*}=\log D+10$, per unit 
absolute magnitude interval for stars with $4<M(g^{'})\leq10$. $<N>$ is 
the mean of number of stars weighted by their numbers. The data are taken from 
Table 10. The other symbols are explained in the text (distances in kpc, volumes 
in pc$^{3}$).}
\begin{tabular}{cccccc}
\hline
$r_{1}-r_{2}$ & $\Delta V_{1,2}$ &      $r^{*}$ &   $z^{*}$ & $<N>$ &  $D^{*}$ \\
\hline
 0.7-1.0  &   4.38 (5) &       0.88 &       0.62 &        191 &       6.64 \\
 1.0-1.5  &   1.58 (6) &       1.30 &       0.92 &        249 &       6.20 \\
 1.5-2.0  &   3.09 (6) &       1.78 &       1.26 &        223 &       5.86 \\
 2.0-2.5  &   5.09 (6) &       2.28 &       1.61 &        181 &       5.55 \\
 2.5-3.0  &   7.59 (6) &       2.77 &       1.96 &        170 &       5.35 \\
 3.0-4.0  &   2.47 (7) &       3.57 &       2.52 &        221 &       4.95 \\
 4.0-5.0  &   4.07 (7) &       4.56 &       3.22 &        157 &       4.59 \\
 5.0-7.5  &   1.98 (8) &       6.49 &       4.58 &        212 &       4.03 \\
 7.5-10.0 &   3.86 (8) &       8.92 &       6.30 &        140 &       3.56 \\
10.0-12.5 &   6.36 (8) &      11.39 &       8.04 &         90 &       3.15 \\
12.5-15.0 &   9.49 (8) &      13.86 &       9.78 &         85 &       2.95 \\
15.0-17.5 &   1.32 (9) &      16.35 &      11.54 &         41 &       2.49 \\
\hline
\end{tabular}  
\end{table}

\begin{table}
\center
\caption{The logarithmic space density function, $D^{*}=\log D+10$, per 
unit absolute magnitude interval for stars with $4<M(g^{'})\leq13$ 
(symbols as in Table 11).}
\begin{tabular}{ccccccc}
\hline
$r_{1}-r_{2}$ & $\Delta V_{1,2}$ &      $r^{*}$   &  $z^{*}$ &  $<N>$ &  $D^{*}$ \\
\hline
 0.2-0.4 &   3.74 (4) &       0.33 &       0.23 &        152 &       7.61 \\
 0.4-0.7 &   1.86 (5) &       0.59 &       0.42 &        293 &       7.20 \\
 0.7-1.0 &   4.38 (5) &       0.88 &       0.62 &        221 &       6.70 \\
 1.0-1.5 &   1.58 (6) &       1.30 &       0.92 &        230 &       6.16 \\
 1.5-2.0 &   3.09 (6) &       1.78 &       1.26 &        223 &       5.86 \\
 2.0-2.5 &   5.09 (6) &       2.28 &       1.61 &        180 &       5.55 \\
 2.5-3.0 &   7.59 (6) &       2.77 &       1.96 &        170 &       5.35 \\
 3.0-4.0 &   2.47 (7) &       3.57 &       2.52 &        277 &       5.05 \\
 4.0-5.0 &   4.07 (7) &       4.56 &       3.22 &        213 &       4.72 \\
 5.0-7.5 &   1.98 (8) &       6.49 &       4.58 &        212 &       4.03 \\
 7.5-10.0&   3.86 (8) &       8.92 &       6.30 &        140 &       3.56 \\
10.0-12.5&   6.36 (8) &      11.39 &       8.04 &         90 &       3.15 \\
12.5-15.0&   9.49 (8) &      13.86 &       9.78 &         85 &       2.95 \\
15.0-17.5&   1.32 (9) &      16.35 &      11.54 &         41 &       2.49 \\
\hline
\end{tabular}  
\end{table}

\begin{table}
\center
\caption{Galactic model parameters estimated by comparison of the 
logarithmic space density function for stars with $4<M(g^{'})\leq10$ (given 
in Table 11) and the combined density laws (Fig. 13). Symbols: 
$n^{*}$ is the logarithmic space density, $H$ is the scaleheight, $n/n_{1}$ 
is the local space density relative to the thin disc and $\kappa$ is the 
axial ratio for the halo.}

\begin{tabular}{cccc}
\hline
Parameter & Thin disc & Thick disc & Halo\\
\hline
$n^{*}$   & $7.51^{+0.04}_{-0.03}$ & $6.32^{+0.06}_{-0.07}$  & $4.41^{+0.38}_{-0.70}$ \\
$H$ (pc)  & $269^{+8}_{-8}$ & $760^{+62}_{-55}$ & $-$ \\
$n/n_{1}$ (per cent) & $-$ & $6.46$ & $0.08$ \\
$\kappa$  & $-$ & $-$& $0.55^{+0.25}_{-0.15}$\\
\hline
\end{tabular}  
\end{table}

\begin{table}
\center
\caption{Galactic model parameters estimated by comparison of the 
logarithmic space density function for stars with $4<M(g^{'})\leq13$ 
(given in Table 12) and the combined density laws (Fig. 14). Symbols 
are as in Table 13.}
\begin{tabular}{cccc}
\hline
Parameter & Thin disc & Thick disc & Halo\\
\hline
$n^{*}$   & $8.18^{+0.01}_{-0.01}$ & $6.36^{+0.08}_{-0.08}$ & $4.41^{+0.56}_{-0.90}$\\
$H$ (pc)  & $173^{+3}_{-3}$       & $822^{+99}_{-85}$   & $-$ \\
$n/n_{1}$ (per cent)  & $-$ & $1.51$ & $0.02$ \\
$\kappa$  & $-$ & $-$ & $0.55^{+0.25}_{-0.15}$\\
\hline
\end{tabular}  
\end{table}

\subsubsection {Model parameters by means of absolute magnitudes 
$4<M(g^{'})\leq13$}

We carried out the work cited in previous paragraph for stars with a 
larger range of absolute magnitude, i.e. $4<M(g^{'})\leq13$. The derived 
density function is given in Table 12 and its comparison with the combined 
density law is shown in Fig. 15. Most of the derived parameters (Table 14), 
especially the local densities, are rather different than the ones cited in 
Section 5.1 and 5.2.1. The reason for this discrepancy is that stars with 
absolute magnitudes $10<M(g^{'})\leq13$ have relatively larger local space 
densities ($Hipparcos$; Jahreiss \& Wielen\ 1997) and are closer to the Sun 
relative to stars brighter than $M(g^{'})=10$, and they affect the combined 
density function considerably. Also the corresponding luminosity function is 
not in agreement with the one of $Hipparcos$ (Fig. 16).

\begin{figure}
\center
\resizebox{8cm}{5.5cm}{\includegraphics*{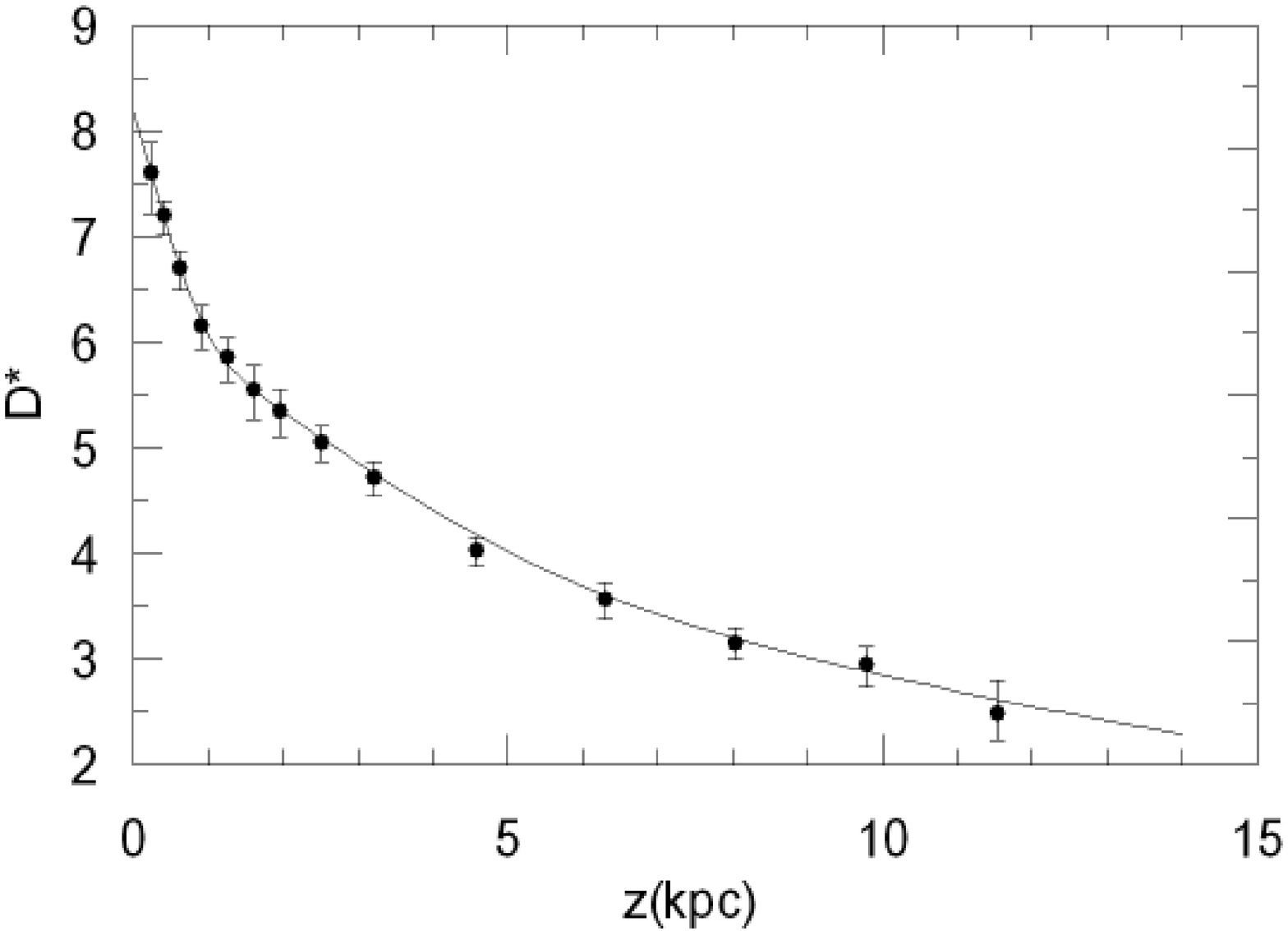}}
\caption[] {Comparison of the derived and combined space density function for 
the thin and thick discs and the halo with the combined density laws, for stars 
with $4<M(g^{'})\leq13$.}
\end{figure}

\begin{figure}
\center
\resizebox{8cm}{5cm}{\includegraphics*{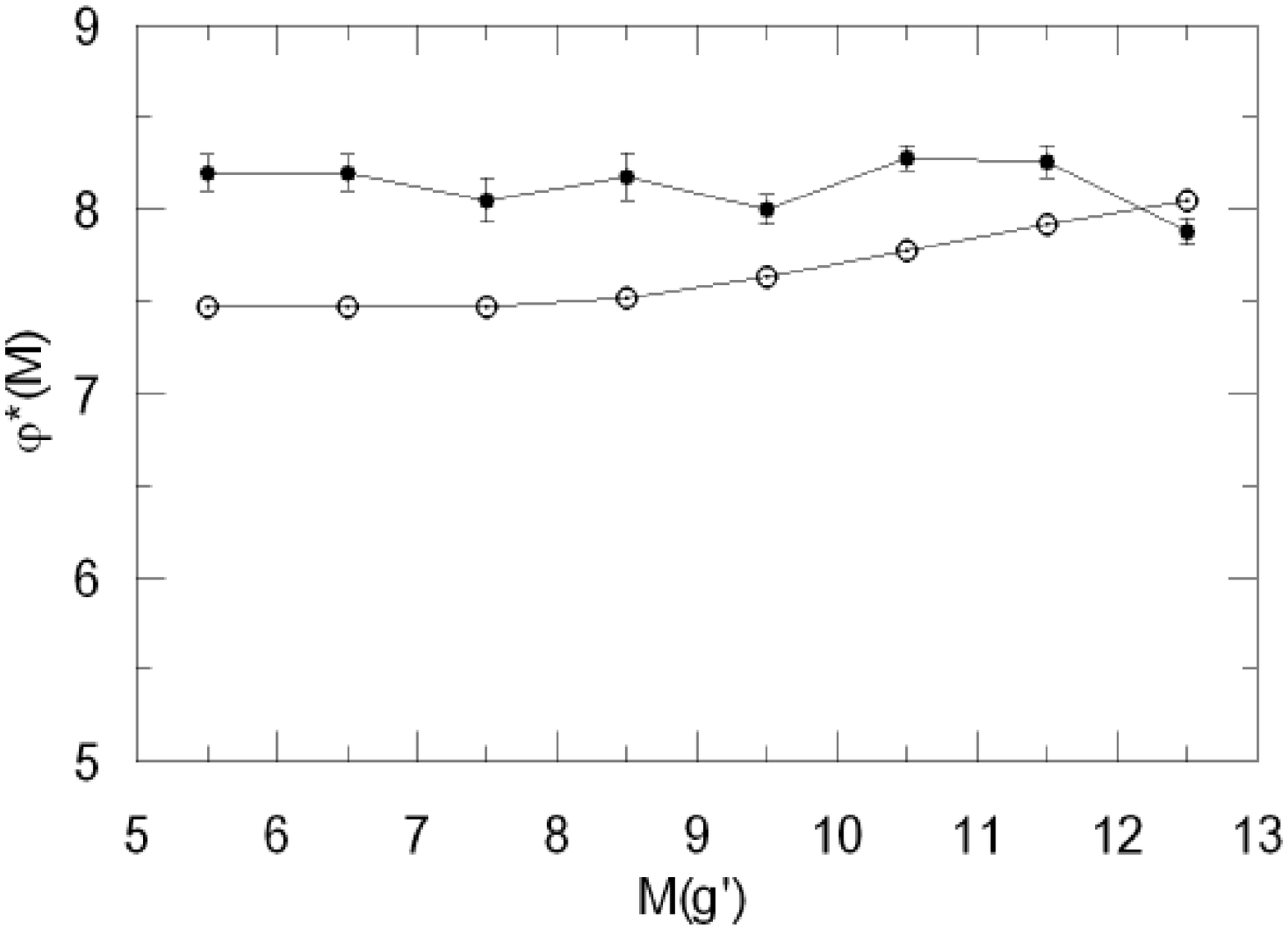}}
\caption[] {The local luminosity function resulting from the comparison of the 
combined derived space density function with the combined density law, for stars with  
$4<M(g^{'})\leq13$.}
\end{figure}

\section {Discussion}

We estimated the Galactic model parameters by comparison of the derived space 
density functions per absolute magnitude interval, in the perpendicular 
direction to the Galactic plane, with a unique density law for each population 
individually for the ELAIS field ($\alpha=16^{h}10^{m}00^{s}, 
\delta= +54^{o}30^{'}00{''}$; $l=84^{o}.27$, $b=+44^{o}.90$; 6.571 deg$^{2}$; 
epoch 2000), by Vega photometry. The separation of stars into different 
populations has been carried out by their spatial position as a function of 
both absolute and apparent magnitude (KBH, see also Karaali\ 1994). This work 
covers nine absolute magnitude intervals, i.e. $4<M(g^{'})\leq5$, 
$5<M(g^{'})\leq6$, $6<M(g^{'})\leq7$, $7<M(g^{'})\leq8$, $8<M(g^{'})\leq9$, 
$9<M(g^{'})\leq10$, $10<M(g^{'})\leq11$, $11<M(g^{'})\leq12$ and 
$12<M(g^{'})\leq13$. However, the populations are not dominant in all absolute 
magnitude intervals. We consider two density laws for the thin-disc stars:
the density law sech$^{2}$ fits better with the derived space density functions 
for the absolute magnitude intervals $10<M(g^{'})\leq11$ and $11<M(g^{'})\leq12$, 
whereas for the absolute magnitude intervals $5<M(g^{'})\leq6$, $6<M(g^{'})\leq7$, 
$7<M(g^{'})\leq8$ and $8<M(g^{'})\leq9$ the exponential density law is 
favorable as in our previous paper (KBH). Contrary to our expectation, for 
the absolutely faintest interval $12<M(g^{'})\leq13$ the exponential law (not 
the sech$^{2}$ one) fits better with the derived space density functions. 
$9<M(g^{'})\leq10$ is a transition absolute magnitude interval where sech fits 
better, a density law which does not correspond to anything physical. The 
scaleheight for the thin disc increases monotonically from 163 to 363 pc, when 
one goes from the absolute magnitude interval $12<M(g^{'})\leq13$ to 
$5<M(g^{'})\leq6$ with the exception scaleheight H=211 pc for the transition 
interval $9<M(g^{'})\leq10$ which is less than the one for the interval 
$10<M(g^{'})\leq11$, i.e. H=295 pc. Some researchers restrict their works 
related with the Galactic model estimation to absolute magnitude. The recent 
work of Robin et al.\ (2003) who treated stars with $M(V)\leq8$ can be given 
as an example. Thus, if we consider the range of the scaleheight only for stars 
with $4<M(g^{'})\leq10$ in our work, we notice that it almost overlaps with the 
range of the scaleheight defined by the minimum and maximum scaleheights given 
in Table 1, i.e. 200-350 pc (we did not take into account the upper limit for 
the thin disc of Robin \& Cr\'{e}z\'{e}\ (1986), 475 pc). The local space 
density for the thin disc decreases monotonically within the uncertainties,  
from absolutely faint magnitude intervals to the bright ones, however, the 
gradient converges to zero at the bright intervals in agreement with the 
corresponding local densities of $Hipparcos$ (Jahreiss \& Wielen\ 1997).

\begin{table*}
\center
\caption{Comparison of the most appropriate Galactic model parameters 
for two works: (SA 114) for our previous work (KBH) and (ELAIS) for 
the present work.}
{\scriptsize  
\begin{tabular}{ccccccccccccc}
\hline
 & \multicolumn{4}{c}{Thin disc} & \multicolumn{4}{c}{Thick disc} & \multicolumn{4}{c}{Halo}\\ 
$M(g^{'})$ & \multicolumn{2}{c}{H (kpc)} & \multicolumn{2}{c}{($n^{*}_{1}$)} & \multicolumn{2}{c}{H (kpc)} 
& \multicolumn{2}{c}{$n_{2}/n_{1}$(per cent)} & \multicolumn{2}{c}{$\kappa$} & \multicolumn{2}{c}{$n_{3}/n_{1}$(per cent)}\\
\hline
Field$\rightarrow$  & SA 114 & ELAIS & SA 114 & ELAIS & SA114 & ELAIS & SA114 & ELAIS& SA 114& ELAIS & SA 114 & ELAIS\\
(4,5]	& & 	    &    &      &    & 0.84 &     &     &      & 0.6 &     &      \\
(5,6]	&0.34& 0.36 & 7.4& 7.4	&0.88& 0.84 & 9.5 & 7.6 &  0.6 & 0.7 & 0.15& 0.06 \\
(6,7]	&0.33& 0.35 & 7.4& 7.4	&0.90& 0.85 & 9.8 & 7.4 &  0.7 & 0.8 & 0.05& 0.04 \\
(7,8]	&0.31& 0.32 & 7.5& 7.4	&0.81& 0.87 & 6.5 & 3.3 &  0.8 &     & 0.02&      \\
(8,9]	&0.29& 0.31 & 7.5& 7.6	&0.97&      & 5.2 &     &      &     &	   &      \\	
(9,10]	&0.26& 0.21 & 7.6& 7.5	&    &      &	  &     &      &     &	   &      \\
(10,11]	&0.30& 0.30 & 8.0& 8.0	&    &      &	  &     &      &     &	   &      \\	
(11,12]	&0.19& 0.26 & 8.6& 8.0	&    &      &	  &     &      &     &	   &      \\	
(12,13]	&0.17& 0.16 & 8.1& 8.0	&    &      &	  &     &      &     &     &      \\
\hline
\end{tabular}
}  
\end{table*}

The logarithmic space densities for the thick disc could be derived for the 
absolute magnitude intervals $4<M(g^{'})\leq5$, $5<M(g^{'})\leq6$, 
$6<M(g^{'})\leq7$ and $7<M(g^{'})\leq8$. The range for the scaleheight is 
rather small, 839-867 pc, and the scaleheight itself is flat within the quoted 
uncertainties. The scaleheight for the thick disc in our work is in good agreement 
with the scaleheight claimed by many authors (cf. Yamagata \& Yoshii\ 1992, von Hippel 
\& Bothun\ 1993, Buser et al.\ 1998; 1999). However, contrary to our expectation, 
the range of the scaleheight for the thick disc cited very recently is large: 
actually Chen et al.\ (2001) and Siegel et al.\ (2002) give 0.58-0.75 and 
0.7-1.0 kpc respectively. Now, let us compare these ranges with the ones in Tables 
13 and 14, where the Galactic model parameters estimated by simultaneous comparison 
to the Galactic stellar populations correspond to stars with $4<M(g^{'})\leq10$ 
and $4<M(g^{'})\leq13$ respectively. The range for the scaleheight of thick disc 
is 705-822 for Table 13 and 737-921 pc for Table 14. We can easily see that they 
are in good agreement, especially if we round our values, i.e. 0.7-0.9 kpc, they 
overlap with the ones of Siegel et al.\ (2002). We should remind that the absolute 
magnitudes of stars treated by Siegel et al.\ (2002) extend down to $M(R)=10.2$ 
which corresponds to $M(g^{'})>10.2$ mag in the Vega photometry. This comparison 
encourage us to argue two points: 

\begin{enumerate}
  \item The absolutely faint stars cause large ranges in the estimation of 
Galactic model  parameters;  
  \item Galactic model parameters are mass (and hence absolute magnitude) 
dependent as claimed in our previous paper (KBH). The local space density 
for the thick disc relative to thin disc increases from 3.31 to 7.59 per cent, 
from absolutely faint magnitude intervals to the bright ones. The large values 
correspond to absolute magnitude intervals where thick disc is dominant (Table 7, 
Fig. 8). Here we reveal another property: although the range of absolute magnitude 
for the thick disc stars, $4<M(g^{'})\leq8$, is not large, the local space density 
for the thick disc cover almost all the range cited in the literature (see 
Table 1) and the larger number of stars results the larger local space density. 
Hence, we can add another point; and
  \item The largeness of the local space density for an absolute magnitude 
interval for a specific population is proportional to the dominance of that 
population in the interval considered. Therefore, if a population is dominant in 
an absolute magnitude interval, then the local space density for that population 
is large. 
\end{enumerate}

The logaritmic space density functions for the halo could be derived only for 
three absolute magnitude intervals, $4<M(g^{'})\leq5$, $5<M(g^{'})\leq6$ and 
$6<M(g^{'})\leq7$, and they are compared with the de Vauculeurs density law. 
The local space density relative to the thin disc $(n_{3}/n_{1})$ could not be 
given for the interval $4<M(g^{'})\leq5$, due to the lack of local space density 
for this interval for the thin disc. For the intervals $5<M(g^{'})\leq6$ and 
$6<M(g^{'})\leq7$, $n_{3}/n_{1}=0.06$ and $n_{3}/n_{1}=0.04$ respectively, 
consistent with the results of Buser et al.\ (1998, 1999) and KBH. The numerical 
values for the axial ratio $\kappa$ for two intervals are close to each other, 
i.e. $\kappa=0.73$ and $\kappa=0.78$ for $5<M(g^{'})\leq6$ and $6<M(g^{'})\leq7$ 
respectively, but a bit less for the interval $4<M(g^{'})\leq5$, $\kappa=0.60$, 
consistent with the previous ones within the uncertainties however. It is 
interesting that the axial ratio estimated by simultaneous comparison to the 
Galactic stellar populations for stars within $4<M(g^{'})\leq13$ is lesser, 
$\kappa=0.55$ (Tables 13 and 14) than the ones cited above. It seems that 
absolutely faint magnitudes where halo stars are very rare causes bad $\kappa$ 
estimation.  

Finally, we compared the Galactic model parameters estimated in our previous 
work ($l=68^{o}.15$, $b=-48^{o}.38$) and in this work ($l=84^{o}.27$, 
$b=+44^{o}.90$) in Table 15. The results confirm the idea that the Galactic 
model parameters are mass (and hence absolute magnitude) dependent. There is a 
good agreement between two sets of data, however, we should note some points 
and keep in mind in the comparison with the results that would appear in future: 
     
~~~(a) Although the scaleheights for a specific absolute magnitude interval of thin 
disc for two fields are rather close to each other, the local space density 
for the interval $11<M(g^{'})\leq12$ is a bit larger for the field SA 114 
($n_{1}^{*}=8.6$) than the one for the ELAIS field ($n_{1}^{*}=8.0$). This slight 
discrepancy probably originates from the previous work because the corresponding 
total local space density does not agree with the $Hipparcos$ (Jahreiss \& Wielen\ 
1997) one either;       

~~~(b) Although the scaleheights estimated for the thick disc for two fields are 
rather close to each other, the local space density relative to the local space 
density of thin disc for the field SA 114 is larger than the one for ELAIS field;
and

~~~(c) The axial ratio, $\kappa$, for the halo for two fields are almost the same, 
whereas the local space density relative the local space density of the thin disc 
for the absolute magnitude interval $5<M(g^{'})\leq6$ for the field SA 114 is 2.5 
times that of the corresponding one for ELAIS field.

\section*{Acknowledgments}
We would like to thank the anonymous referee for insightful comments and suggestions 
that helped to improved this paper. 
We wish to thank all those who participated in observations of the ELAIS field. 
The data were obtained through the Isaac Newton Group's Wide Field Camera Survey 
Programme, where the Isaac Newton Telescope is operated on the island of La Palma 
by the Isaac Newton Group in the Spanish Observatorio del Roque de los Muchashos 
of the Instituto de Astrofisica de Canaries. We also thank CASU for their data 
reduction and astrometric calibrations.This work was supported by the Research 
Fund of the University of Istanbul. Project number: 1417/05052000.

\end{document}